\documentclass{article}
\usepackage{arxiv}
\usepackage[utf8]{inputenc}
\usepackage[T1]{fontenc}
\usepackage{hyperref}
\usepackage{url}
\usepackage{booktabs}
\usepackage{amsfonts}
\usepackage{nicefrac}
\usepackage{microtype}
\usepackage{amsmath,amsthm,amssymb,bm,mathtools}
\usepackage{verbatim}
\usepackage{graphicx}
\usepackage{xcolor}
\usepackage[normalem]{ulem}
\usepackage{subcaption}
\usepackage{algorithm,algpseudocode}
\usepackage{enumitem}
\usepackage{multirow}
\usepackage{array}
\usepackage{siunitx}
\usepackage{float}
\usepackage[section,below]{placeins}
\usepackage[numbers]{natbib}
\usepackage[capitalize,nameinlink]{cleveref}

\DeclareMathOperator{\diag}{diag}

\newcommand{\kap}{\boldsymbol{\kappa}}

\newcommand{\bu}{\bm{u}}

\newcommand{\bd}{\mathbf{d}}

\title{Latent Diffusion Posterior Sampling with Surrogate Likelihood Guidance for PDE Inverse Problems}

\author{
Yuanzhe Wang \\ Department of Civil and Environmental Engineering \\ University of Illinois Urbana-Champaign \\ Urbana, IL 61801 \\ \texttt{yuanzhe9@illinois.edu} \\ \And Alexandre M.\ Tartakovsky \\ Department of Civil and Environmental Engineering \\ University of Illinois Urbana-Champaign \\ Urbana, IL 61801 \\ \texttt{amt1998@illinois.edu} }

\begin{document}

\maketitle

\begin{abstract}

We propose latent-space diffusion posterior sampling (L-DPS), an approximate Bayesian framework for high-dimensional inverse problems governed by partial differential equations (PDEs). The method is motivated by three central challenges in PDE-constrained inversion: 
the absence of a tractable explicit form of the prior distribution when the prior is specified only through samples, 
the high dimensionality of spatially distributed parameters, 
and the high cost of repeated forward-model evaluations during posterior sampling. L-DPS addresses these challenges by combining a variational autoencoder (VAE), an unconditional latent diffusion model, diffusion posterior sampling (DPS), and a differentiable neural surrogate. The VAE maps the parameter field to a lower-dimensional latent feature space, the unconditional diffusion model learns an implicit prior score in this latent space, and DPS combines this learned prior with likelihood-based guidance. The likelihood gradient is evaluated through the decoder--surrogate composition, avoiding repeated calls to the full numerical PDE solver.

We evaluate L-DPS on an inverse Darcy flow problem with an unknown spatially distributed permeability field inferred from sparse and noisy pressure observations. The proposed method produces accurate and robust inverse solutions. It reduces the inference cost relative to full-space DPS while outperforming amortized inverse baselines such as conditional latent diffusion and inverse FNO in sparse and noisy regimes. We also compare L-DPS with a KLE-MAP baseline, in which the parameter field is represented by a Karhunen--Lo\`eve expansion and the latent prior is approximated as Gaussian. L-DPS gives more accurate inverse solutions for the Gaussian priors considered here, and its advantage increases for non-Gaussian parameter fields, where the Gaussian approximation underlying KLE-MAP becomes less effective.

We further investigate whether a single mixed-prior, or foundational, L-DPS model can be used without specifying the parameter prior at inference time. The foundational model retains competitive accuracy across Gaussian and binary prior classes, although prior-specific L-DPS models remain more accurate when the test prior is known. Finally, we quantify the sensitivity of L-DPS to surrogate model accuracy using the Fourier Neural Operator (FNO), Vision Transformer, and Deep Operator Network (DeepONet). The inverse error increases monotonically with surrogate forward error, indicating that surrogate accuracy, rather than surrogate architecture alone, is a key factor controlling the quality of L-DPS inversion.

\end{abstract}


\section{Introduction}\label{sec:intro}

We propose a latent-space diffusion posterior sampling method (L-DPS), an approximate Bayesian method for solving high-dimensional inverse partial differential problems in which the functional form of the (non-Gaussian) prior distribution (density) of the unknown parameters is unavailable, and only samples from the prior are given.  
Bayesian inverse problems require sampling from the posterior distribution $p(y \mid d)$,
where the choice and representation of the prior $p(z)$ play a central role in determining the appropriate inference methodology. 
The methods for posterior sampling can be classified into the following three categories, depending on the type of access they require to the prior distribution: 

(i) Density-based methods require evaluation of the prior density $p(y)$ or its logarithm $-\log p(y)$, and include approaches such as maximum a posteriori (MAP) estimation, randomized MAP, and Markov Chain Monte Carlo (MCMC) methods, including Metropolis--Hastings and Hamiltonian Monte Carlo. 

(ii) Score-based methods require evaluation of the prior score $\nabla_y \log p(y)$, and include diffusion posterior sampling (DPS), score-based Langevin dynamics, and sampling methods based on stochastic differential equations.

(iii) Sample-based methods require only the ability to draw samples from the prior distribution, without explicit access to its density or score. These include likelihood-free inference methods such as approximate Bayesian computation (ABC), as well as simulation-based approaches and particle methods in settings where only sampling from the prior or forward model is available.

When the prior distribution is \emph{explicitly known}, i.e., when its density $p(z)$ or its logarithm is available in closed form, all of the aforementioned methods can, in principle, be applied. In many practical applications, however, the parameter $z$ is (1) high-dimensional, and (2) the prior may only be accessible through samples or an implicit generative model. These two challenges significantly limit the applicability of standard approaches.

For the first challenge, many classical sampling methods, including Markov chain Monte Carlo (MCMC) and sample-based methods, suffer from the curse of dimensionality, with computational cost and convergence rates deteriorating rapidly as the dimension of $z$ increases. For the second challenge, density-based and score-based methods require access to the prior density $p(z)$ or its gradient $\nabla_z \log p(z)$, which may not be available when the prior is specified implicitly through data or complex generative processes.

Addressing these two challenges typically requires additional tools, such as dimensionality reduction, surrogate modeling, or learning-based approaches that approximate the prior distribution or its score function. These considerations motivate the use of latent-variable models and diffusion-based methods, which provide tractable representations of complex high-dimensional priors while enabling efficient posterior sampling.

Both challenges can be potentially removed using the expansions of the complex high-dimensional distributions in terms of a relatively small number of random variables $\bm{z}$. For example, the Karhunen--Lo\`eve expansion (KLE) uses a linear approximation of $y$. When $y$ is a multi-dimensional Gaussian, KLE can represent it as a linear combination of independent Gaussian variables $\bm{z}$. The variational autoencoders (VAE) use a non-linear combination of $\bm{z}$ to represent a non-Gaussian $y$. While $\bm{z}$ is regularized toward an independent standard Normal distribution $\mathcal{N}(0,I)$, good approximation properties of VAE usually require the regularization to be weak, yielding the distribution of $\bm{z}$ to deviate from $\mathcal{N}(0,I)$ and be given implicitly by samples. 
In summary, for a complex non-Gaussian $y$, its distribution or the distribution of $\bm{z}$ in its expansion, is not explicitly known, which limits the choice of the sampling methods to the likelihood-free sampling methods, which suffer from the CoD.     

One alternative is to approximate the non-Gaussian distribution of $y$ or its latent variables with a Gaussian distribution, whose mean and variance are computed from the samples of $y$ or $z$. Then, both likelihood-based and score-based methods can be applied; however, the accuracy of the posterior estimate can be severely affected by how far the true prior distributions are from the Gaussian distribution. Another alternative solution is to use the Denoising Diffusion Probabilistic Model (DDPM) \cite{song2021score}, which approximates the prior score using a denoising network.
It is important to note that $p(y)$ or $\log p(y)$ cannot be explicitly obtained from DDPM, which limits sampling to the score-based methods. 
The DDPM was used to model the prior score of the high-dimensional functions as well as their latent variables $z$. Of particular relevance to our work is the latent diffusion model (LDM) \cite{rombach2022high}, in which the DDPM was applied to explicitly model the prior score of $z$ defined by the VAE. 

 DDPMs have mostly been used for generative purposes. DDPMs were also used to sample Bayesian posteriors for image restoration, enhancement, and inverse imaging. Of particular interest to our work is the diffusion posterior sampling (DPS) method \cite{chung2023diffusion}, which uses the prior score provided by a DDPM to sample the posterior via an iterative scheme, in which the prior score of $y$ is combined with a likelihood-based correction to guide sampling toward the posterior distribution. In this work, the DPS was used to deblur images; the DDPM was used to learn the prior score of $y$ from the pixel map; and the likelihood $p(d \mid y)$ was computed using the blurring operator. In \cite{feng2025surgin}, a differentiable surrogate model was combined with a DPS-type score-based generative inversion for subsurface flow problems.

DPS direct application to high-dimensional PDE inverse problems is challenging because posterior guidance requires repeated evaluations of the forward model and its gradients \cite{ho2020denoising,chung2022diffusion,kawar2022denoising}.  In our work, we propose DPS sampling in the latent space for high-dimensional inverse PDE problems. Our method combines the LDM and DPS methods and uses a surrogate neural operator model to compute the likelihood. Its novelty lies in extending the DPS method to physics-constrained high-dimensional inverse problems by learning a DDPM in latent space and imposing physics constraints via a neural operator. Our numerical examples demonstrate that both latent representation (as opposed to learning in the space of $y$) and a non-Gaussian prior score model (as opposed to the Gaussian approximation) are important for obtaining accurate inverse solutions.  

The summary of Bayesian methods, provided above, is not intended to be exhaustive. Most notably, it does not discuss likelihood-free methods, including the normalizing flows and conditional diffusion models. 
Our work is focused on posterior sampling methods that explicitly incorporate the likelihood function and are therefore applicable to general inverse problems defined by a forward operator. The normalizing flows and conditional diffusion models aim to learn a direct mapping from observations $d$ to latent variables $z$ or parameters $y$, effectively approximating the conditional distribution $p(z \mid d)$ (or $p(y \mid d)$) through supervised or amortized training. While such approaches can provide fast inference once trained, they require large datasets of paired samples $(z,d)$, are typically tied to a fixed observation operator and noise model, and may suffer from generalization issues when applied to new measurement configurations.

In contrast, diffusion posterior sampling (DPS) operates in a fundamentally different regime. Rather than learning the posterior directly, DPS combines a learned prior model, represented by a diffusion process, with an explicit likelihood term derived from the forward model. This allows DPS to be applied to new observation settings without retraining, accommodates arbitrary forward operators and noise levels, and provides a mechanism for incorporating physical models directly into the inference process. Furthermore, DPS relies on score-based sampling and thus leverages the full structure of the learned prior distribution without requiring paired training data. These distinctions make DPS particularly well-suited for physics-constrained inverse problems, where forward models are known but generating representative supervised datasets for all possible observation scenarios is infeasible. In this study, L-DPS is used as a posterior-guided single-sample reconstruction method; uncertainty quantification by drawing larger ensembles is left for future work.

The main contributions of this work are:
(i) a latent-space DPS formulation for high-dimensional PDE inverse problems with sample-based, implicit priors;
(ii) a surrogate-likelihood guidance strategy that differentiates through the decoder--surrogate composition and avoids repeated PDE solves;
(iii) a normalized, noise- and observation-density-adaptive guidance rule for stabilizing posterior-guided reverse diffusion; and
(iv) a systematic numerical study of latent compression, surrogate accuracy, guidance stabilization, Gaussian versus non-Gaussian priors, and mixed-prior generalization.

This paper is organized as follows. The forward and inverse PDE problems are formulated in Section \ref{sec:general_inverse}. Section \ref{sec:method} provides background information on the latent space unconditional diffusion model. The L-DPS model is formulated in Section \ref{sec:dps_latent}. Section \ref{sec:darcy_case} presents the numerical results and ablation study. Conclusions are given in Section \ref{sec:conclusion}.

\section{Problem Formulation: General PDE Inverse Problem}\label{sec:general_inverse}

Consider a partial differential equation (PDE) problem of the form:
\begin{align}\label{eq:general_pde}
  \mathcal{L}(u(\bm{x});\,\kappa(\bm{x})) = 0, \quad \bm{x} \in \Omega,
\end{align}
subject to appropriate boundary conditions and forcing terms. Here $\mathcal{L}$ is a differential operator, $\kappa(\bm{x})$ is the unknown space-dependent parameter field, $u(\bm{x})$ is the state variable, and $\Omega$ is the computational domain. 

The inverse problem is to estimate $\kappa(\bm{x})$ from sparse and noisy observations $\mathbf{d} = [d_1,...,d_{N_o}]^T$ of the system response $u(\bm{x})$ at the locations $\bm{x}_o = [\bm{x}_{o,1},...,\bm{x}_{o,N_o}]^T$. In general, the function $\kappa(\bm{x})$ is infinite-dimensional and cannot be estimated exactly. Therefore, it is common to discretize the domain $\Omega$ with $N$ elements as part of the numerical solution of the PDE problem and replace $\kappa(\bm{x})$ with the $N$-dimensional vector $\kap$ of the parameter field defined at each element. Then the inverse problem reduces to estimating $\kap$. This inverse problem is ill-posed because, in general, the number of observations $N_o$ is much smaller than $N$. 

Following a standard Bayesian approach, we employ the observation model
\begin{align}
  \mathbf{d} \;=\; \mathbf{u}_o(\kap)
        \;+\; \gamma\,\frac{\|\mathbf{u}_o(\kap)\|_2}{\sqrt{N_{\text{obs}}}}\,\boldsymbol{\epsilon}_0,
  \qquad \boldsymbol{\epsilon}_0 \sim \mathcal{N}(\mathbf{0}, \mathbf{I}),
  \label{eq:obs}
\end{align}
where $\mathbf{u}_o(\kap) = [u(\bm{x}_{o,1},\kap),\ldots,u(\bm{x}_{o,N_{\text{obs}}},\kap)]^T$ is the vector of the PDE solution at the sensor locations and $\gamma \ge 0$ is a dimensionless noise level (coefficient of variation) independent of the number of observations. 

We adopt the Bayesian formulation of the inverse problem, in which the parameter field $\kap$ is treated as a random field with prior distribution $p(\kap)$. Then, the posterior distribution of $\kap$ is given by the Bayes formula 
\begin{align}\label{eq:posterior}
  p(\kap \mid \mathbf{d}) \propto
  p(\mathbf{d} \mid \kap)\,p(\kap),
\end{align}
where the likelihood $p(\bm{d}\mid\kap)$ model corresponding to the observation model in Eq \eqref{eq:obs} has the form
\begin{align}\label{eq:likelihood}
  p(\mathbf{d}\mid\kap) \;\propto\;
  \exp\!\Big(-\tfrac{1}{2\sigma^2}\,\|\mathbf{d}-{\mathbf{u}}_o(\kap)\|_2^2\Big),
  \qquad \sigma \;=\; \gamma\,\frac{\|\mathbf{d}\|_2}{\sqrt{N_{\text{obs}}}}.
\end{align}

The posterior distribution provides a probabilistic description of the inverse solution accounting for uncertainty in the parameter field induced by limited and noisy observations. We assume that the explicit form of $p(\kap)$ is unknown and only samples of $\kap$ are available, as is the case for most natural systems. We develop a latent-space DPS framework to sample $p(\kap \mid \mathbf{d})$. This approach first reduces the dimensionality of $\kap$ to that of the latent variables $\bm{z}$ as defined by a VAE. Then it approximates the score of the prior distribution of $\bm{z}$ using the unconditional diffusion model. Then, the DPS model is used to sample the posterior distribution of $\bm{z}$ using its score. 

Furthermore, we assume that a numerical solution of $\bm{u}$ on the $N$-element mesh (the same mesh that is used to discretize $\kappa(x)$) is too computationally expensive to repeatedly evaluate in the iterative DPS sampling. We approximate the numerical PDE solution with a neural operator model $\mathcal{F}$ mapping $\kap$ to the state variable $u(x)$: 
\begin{align}\label{eq:forward}
  u(x,\kap) \approx  \hat{u}(x,\kap) = \mathcal{F}(x,\kap).
\end{align}
We use $\mathcal{F}(x,\kap)$ to approximate the PDE solution at the measurement locations, which we denote as 
\begin{equation}
\hat{\mathbf{u}}_o(\kap) = [\mathcal{F}(\bm{x}_{o,1},\kap),\ldots,\mathcal{F}(\bm{x}_{o,N_{\text{obs}}},\kap)]^T.
\end{equation}

The neural operator is trained using a labeled data set $D = \{\kap^{(i)} \rightarrow \bu^{(i)} \}_{i=1}^{N_{train}}$ consisting of the $N_{train}$ numerical solutions of the PDE problems for different realizations $\kap$ randomly drawn from a mixture of different distributions. Here, $\bu^{(i)}$ is the numerical PDE solution corresponding to the parameter vector $\kap^{(i)}$.
We consider several forms of neural operators, which are discussed in \cref{sec:darcy_case}.

\section{Background: Latent-Space Unconditional Diffusion Model}\label{sec:method}
\subsection{ Latent Representation of the Parameter Field}\label{sec:ldps_overview}
Sampling high-dimensional posterior distributions is computationally demanding and can yield poor parameter prediction. Therefore, in this work, we propose to perform DPS sampling in the latent space of $\kap$. 
Our numerical examples demonstrate that DPS sampling in the physical space of $\kap$ may yield less accurate predictions than sampling its latent posterior.  

In this work, the latent representation of $\kap$ is obtained with a VAE-like model presented in \cite{rombach2022high}. Let $E_{\psi}$ and $D_{\psi}$  denote the encoder and decoder, respectively, with parameters $\psi$. The VAE maps the parameter vector $\kap$ to a latent variable vector using an encoder 
\begin{align}
    \bm{z} = E_{\psi}(\kap):=\bm\mu_\psi + \bm\sigma_\psi \odot \boldsymbol\epsilon, \quad \boldsymbol\epsilon \sim \mathcal{N}(\mathbf 0,\mathbf I),
\end{align}
where the $\odot$ is the Hadamard product, the mean $\bm\mu_\psi$ and standard deviation $\bm\sigma_\psi$ are the outputs of $E$, and $\bm{z}$ is a random variable with the Gaussian conditional distribution 
\begin{align}
    \bm{z} \sim q_{\psi}(\bm{z} \mid \kap)=\mathcal{N}(\bm\mu_\psi,\Sigma_\psi),
\end{align}
 where $\Sigma_\psi = \diag(\bm\sigma_\psi^2)$ is the diagonal covariance matrix. 
 
The parameter vector is then reconstructed from the latent variable through the decoder as
\begin{align}
    \kap \approx D_{\psi}(\bm{z}).
\end{align}

The VAE is trained to minimize the reconstruction loss together with a regularization term that promotes a structured latent space. Specifically, the training objective is given by
\begin{align}
    \mathcal{L}_{VAE} (\psi) = \mathbb{E}_{q_{\psi}(z \mid \kappa)}
\left[
\left\|\kap - D_{\psi}[E_{\psi}(\kap)]\right\|_2^2
\right]
+
\beta\, D_{\mathrm{KL}}
\!\left(
q_{\psi}(\bm{z} \mid \kap)\,\|\,p_t(\bm{z})
\right),
\end{align}
where $p_t(\bm{z})$ is the target distribution of $\bm{z}$, $D_{KL}$ is the Kullback–Leibler divergence, and $\beta$ is a weighting parameter. The distribution $p_t(\bm{z})$ is assumed to be standard Gaussian  $\mathcal{N}(0,I)$. 
The first term enforces accurate reconstruction of the parameter field. In contrast, the second term regularizes the latent representation and facilitates the subsequent learning of a generative prior in the latent space.

Here, we represent the discretized parameter field as a multi-channel tensor. 
For a three-dimensional scalar parameter field, this tensor has the form
\[
    \bm{\kappa} \in \mathbb{R}^{C_{\kappa}\times H\times W\times D},
\]
where \(H,W,D\) are the numbers of grid cells in the three spatial directions and 
\(C_{\kappa}\) is the number of physical input channels. 
For the scalar coefficient fields considered in this work, \(C_{\kappa}=1\). 
The channel dimension is explicitly included because the encoder and decoder are convolutional networks, and their internal representations are naturally organized into multi-channel feature maps, as in standard CNNs.

The VAE encoder maps the parameter field to a lower-resolution latent feature tensor,
\[
    \bm{z}
    =
    E_{\psi}(\bm{\kappa})
    \in
    \mathbb{R}^{C_z\times H/f\times W/f\times D/f},
\]
where \(f\) is the spatial downsampling factor and \(C_z\) is the number of latent channels. 
The last three dimensions of \(\bm z\) retain the spatial organization of the original field, but on a coarser grid. 
The first dimension, \(C_z\), is not an additional spatial dimension; it is a learned feature-channel dimension. 
Each latent channel stores a different learned representation of the spatial structure of the parameter field, analogous to feature channels in a convolutional neural network.

The downsampling factor \(f\) and the number of latent channels \(C_z\) jointly determine the dimension of the latent space. 
For a scalar three-dimensional field with \(C_{\kappa}=1\), the full parameter dimension is
\[
    N_{\kappa}=HWD,
\]
whereas the latent dimension is
\[
    N_z
    =
    C_z\frac{HWD}{f^3}.
\]
Thus, the compression ratio is
\[
    \frac{N_{\kappa}}{N_z}
    =
    \frac{f^3}{C_z}.
\]

Both \(C_z\) and \(f\) are architectural hyperparameters. 
Increasing \(f\) reduces the spatial resolution of the latent representation, while increasing \(C_z\) increases the number of learned feature maps and can partially compensate for the information lost by spatial downsampling. 
In practice, \(f\) and \(C_z\) are chosen together to obtain a latent representation that is sufficiently compressed for efficient posterior sampling while still accurate enough to reconstruct the parameter field.

Having such multidimensional $\bm{z}$ is different from previous work on latent-space neural operators and image synthesis, which use a one-dimensional, arbitrarily ordered latent space and a relatively large $\beta$, yielding the latent variables to be approximately $\mathcal{N}(0,1)$ \cite{esser2021taming,ramesh2021zero,zong2025vae}. It was shown in \cite{rombach2022high} that a multidimensional correlated latent space in combination with a relatively small $\beta$ better preserves the information of $\kap$ than a one-dimensional uncorrelated latent space and provides a more accurate reconstruction, while the latter provides a better generative model.

Setting a relatively large $\beta$ strongly constrains $q_\psi(z \mid \kappa)$ to match $p_t(z)=\mathcal{N}(\mathbf 0,\mathbf I)$. As a result, the aggregated latent distribution \[q_{\mathrm{agg}}(z)=\int q_\psi(z \mid \kappa)\,p_{\mathrm{data}}(\kappa)\,d\kappa\] is close to a standard Gaussian.  However, this constraint reduces the flexibility of the latent representation and may degrade reconstruction accuracy, since the encoder is restricted to use only Gaussian-compatible features. 

In contrast, in the \emph{weak-KL regime} ($\beta \ll 1$), the reconstruction term dominates the objective, allowing the encoder--decoder pair to learn a highly accurate representation of the parameter field. However, in this regime, $\bm\mu_\psi$ and $\Sigma_\psi$ will deviate from 0 and $I$, respectively, and will be different for different $\kap$ samples. Consequently, the aggregated latent distribution $q_{\mathrm{agg}}(z)$ deviates significantly from $\mathcal{N}(0,I)$ and becomes a complex, structured, and generally non-Gaussian distribution.

This non-Gaussianity is particularly pronounced for spatially correlated parameter fields such as permeability, where the encoder captures nonlinear combinations of spatial patterns, leading to dependencies and multimodal structures in the multi-dimensional latent space.

The aggregated distribution of $\bm{z}$ provides an estimate of the $\bm{z}$ prior distribution, $p(z)\approx q_{agg}(\bm{z})$, which can be used in the Bayesian formula to define the posterior distribution of $\bm{z}$ conditioned on data:
\begin{align}\label{eq:latent_posterior}
  p(\bm{z} \mid \mathbf{d}) \propto
  p(\mathbf{d} \mid \bm{z})\,p(\bm{z}).
\end{align}

Since the size of $\bm{z}$ is less than that of $\kap$, sampling the posterior distribution of $\bm{z}$ is more efficient than sampling the distribution of $\kap$. 

To sample the posterior distribution $p(\bm{z} \mid \mathbf{d})$ using methods such as MCMC, HMC, or the randomized MAP method, we need to know the functional form of $p(\bm{z})$.  

The samples of $p(\bm{z})$ can be obtained as $\bm{z}^{(i)} = E(\kap^{(i)})$. However, these samples are insufficient for methods that require likelihood evaluation.  

Following \cite{rombach2022high}, we use the DDPM ~\cite{ho2020denoising} to obtain a tractable generative representation of $p(\bm{z})$. The DDPM model is trained on samples from $p(\bm{z})$.  Since diffusion models evolve samples in time following the diffusion process, in the following, we denote the VAE latent variable as $\bm{z}_0$, the ``clean'' (i.e., not corrupted by the diffusion noise) variable, and the samples of $p(\bm{z}_0)$ obtained with $E$ from different samples of $\kap^{(i)}$ as $\bm{z}_0^{(i)}$.

\subsection{DDPM model for learning the latent prior score}\label{sec:ldps_prior}

We employ a denoising diffusion probabilistic model (DDPM) to construct a tractable representation of the prior distribution $p(\bm{z}_0)$ from samples $\{\bm{z}_0^{(i)}\}$. Specifically, we use DDPM to approximate the prior score (the gradient of the log of the $\bm{z}_0$ prior distribution) and to sample $\bm{z}_0$. The DDPM consists of a forward (noising) process that gradually transforms $\bm{z}_0$ into a standard Gaussian variable, and a reverse (denoising) process, parameterized by a neural network, that approximately inverts this transformation. In \cref{sec:dps_latent}, this learned prior model is combined with a likelihood-based correction to draw approximate samples from the posterior $p(\bm{z}_0 \mid \bm{d})$.

\paragraph{Forward diffusion process.}
The forward process is defined as a Markov chain with transitions
\begin{align}
    \bm{z}_t
    \;=\;
    \sqrt{\alpha_t}\,\bm{z}_{t-1}
    \;+\;
    \sqrt{1-\alpha_t}\,\bm{\epsilon}_t,
    \qquad
    \bm{\epsilon}_t \sim \mathcal{N}(\bm{0}, \mathbf{I}),
    \qquad t=1,\dots,T,
\end{align}
where $\{\alpha_t\}_{t=1}^T$ is a sequence of noise parameters. By recursively applying this relation, one obtains the closed-form marginal distribution
\begin{align}
    \bm{z}_t
    \;=\;
    \sqrt{\bar{\alpha}_t}\,\bm{z}_0
    \;+\;
    \sqrt{1-\bar{\alpha}_t}\,\bm{\epsilon},
    \qquad
    \bm{\epsilon} \sim \mathcal{N}(\bm{0}, \mathbf{I}),
    \label{eq:forward_marginal}
\end{align}
where $\bar{\alpha}_t = \prod_{s=1}^t \alpha_s$. This expression defines the distribution of $\bm{z}_t$ conditioned on $\bm{z}_0$ without explicitly simulating intermediate steps. As $t \to T$, the distribution of $\bm{z}_t$ approaches $\mathcal{N}(\bm{0}, \mathbf{I})$.

\paragraph{Reverse process and training.}
The reverse process is parameterized by a neural network $\bm{\epsilon}_\omega(\bm{z}_t,t)$ trained to predict the noise component from $(\bm{z}_t,t)$. The parameters $\omega$ are obtained by minimizing
\begin{align}
    \mathcal{L}_{\mathrm{diff}}
    \;=\;
    \mathbb{E}_{\bm{z}_0,\,\bm{\epsilon},\,t}
    \left[
        \left\|
            \bm{\epsilon}
            -
            \bm{\epsilon}_\omega(\bm{z}_t,t)
        \right\|_2^2
    \right],
    \label{eq:diff_loss}
\end{align}
where the expectation is taken over latent samples $\bm{z}_0$, Gaussian noise $\bm{\epsilon}$, and diffusion times $t$.

The relation between the noise predictor and the score follows from the Gaussian
forward marginal in Eq.~\eqref{eq:forward_marginal}. Defining $\sigma_t^2=1-\bar{\alpha}_t$, the distribution of $\bm{z}_t$ conditioned on $\bm{z}_0$ is 
$q(\bm z_t\mid \bm z_0)=
\mathcal N(\sqrt{\bar{\alpha}_t}\bm z_0,\sigma_t^2\mathbf I)$. Therefore, 
\[
    \nabla_{\bm z_t}\log q(\bm z_t\mid \bm z_0)
    =
    -\frac{\bm z_t-\sqrt{\bar{\alpha}_t}\bm z_0}{\sigma_t^2}
    =
    -\frac{1}{\sigma_t}\bm\epsilon .
\]
Averaging over $p(\bm z_0\mid \bm z_t)$ gives
\[
    \nabla_{\bm z_t}\log p_t(\bm z_t)
    =
    -\frac{1}{\sigma_t}
    \mathbb E[\bm\epsilon\mid \bm z_t].
\]
Since the noise-prediction loss in Eq.~\eqref{eq:diff_loss} trains
$\bm\epsilon_\omega(\bm z_t,t)$ to approximate
$\mathbb E[\bm\epsilon\mid \bm z_t]$, the prior score is approximated by
\begin{align}\label{eq:score}
    \nabla_{\bm{z}_t} \log p_t(\bm{z}_t)
    \approx
    -\frac{1}{\sqrt{1-\bar{\alpha}_t}}\,
    \bm{\epsilon}_\omega(\bm{z}_t,t),
\end{align}
which provides the prior score used in the posterior sampling scheme described in \cref{sec:dps_latent}.

It is important to emphasize that the DDPM does not provide an explicit
closed-form density $p(\bm z_0)$ or log-density $\log p(\bm z_0)$.
Instead, the model defines an implicit generative distribution through the
reverse diffusion chain. Evaluating $p(\bm z_0)$ would require marginalizing
over all intermediate latent variables in the reverse process, which is
generally intractable. The quantity made available by the trained DDPM is
therefore not the density itself, but the score of the noisy marginal
distribution $p_t(\bm z_t)$, approximated through the noise-prediction
network as in Eq.~\eqref{eq:score}. This is why the learned prior is used
within a score-based posterior sampler rather than in density-based methods
such as MAP or HMC, requiring explicit evaluation of $\log p(\bm z_0)$.

\paragraph{Sampling and denoising estimate.}
At inference, we use the deterministic denoising diffusion implicit model (DDIM) sampler~\cite{song2020denoising}, obtained by setting the per-step noise variance to zero. Starting from $\bm{z}_T \sim \mathcal{N}(\bm{0}, \mathbf{I})$, the reverse process yields a sample $\bm{z}_0$ approximately distributed according to $p(\bm{z}_0)$.

An estimate of the clean latent variable $\bm{z}_0$ at time $t$ can be obtained as
\begin{align}
    \hat{\bm{z}}_0(\bm{z}_t)
    \;=\;
    \frac{
        \bm{z}_t
        -
        \sqrt{1-\bar{\alpha}_t}\,
        \bm{\epsilon}_\omega(\bm{z}_t,t)
    }{
        \sqrt{\bar{\alpha}_t}
    }.
    \label{eq:hat_z0}
\end{align}
Both the score and $\hat{\bm{z}}_0$ DDPM estimates are used in \cref{sec:dps_latent} to construct tractable approximations of the log posterior gradient for solving the inverse problem.

We distinguish between the trained diffusion model and the sampling scheme. 
The denoising network \(\epsilon_\omega(z_t,t)\) is trained with the DDPM noise-prediction objective and is used to approximate the latent prior score. 
At inference, however, we use the DDIM reverse update as a deterministic sampler for this trained model by setting the per-step sampling variance to zero. 
Thus, DDPM refers to the learned latent diffusion prior, whereas DDIM refers only to the particular reverse sampler used to generate samples from that prior.

\section{Latent-Space Diffusion Posterior Sampling}
\label{sec:dps_latent}

\subsection{L-DPS Model Formulation}
The objective of the online inversion stage is to infer the clean latent variable
\(\bm z_0\), and hence the parameter field \(\kappa = D_\psi(\bm z_0)\), from the observations \(\bm d\). In Bayesian form, the target posterior in latent space is
\begin{align}
    p(\bm z_0 \mid \bm d)
    \propto
    p(\bm d \mid \bm z_0)\,p(\bm z_0),
\end{align}
where \(p(\bm z_0)\) is the latent prior learned by the diffusion model and
\(p(\bm d \mid \bm z_0)\) is the likelihood induced by the decoded parameter field and the forward PDE model.

Standard diffusion posterior sampling (DPS) solves inverse problems by combining an unconditional diffusion model, which provides an approximation to the prior score, with a likelihood-gradient correction that enforces agreement with observations during the reverse diffusion process~\cite{chung2023diffusion}. The original DPS formulation was developed for image inverse problems, where the unknown is typically an image, the reverse diffusion process is carried out in the ambient image space, and the likelihood gradient is evaluated through a prescribed measurement operator. In contrast, we perform posterior-guided sampling in the VAE latent space and evaluate the likelihood gradient through the decoder and the differentiable neural surrogate.

The proposed latent-space DPS (L-DPS) model is summarized in \cref{alg:ldps_compact}. The algorithm starts with a Gaussian latent variable \(\bm z_T \sim \mathcal N(0,I)\). At each reverse diffusion step \(t\), the diffusion model predicts the noise component \(\hat{\bm\epsilon}_t = \bm\epsilon_\omega(\bm z_t,t)\). This prediction is used in two ways. First, it defines the unconditional reverse diffusion update \(\bm z'_{t-1}\), which moves the sample according to the learned latent prior. Second, it gives the Tweedie denoised estimate \(\hat{\bm z}_0(\bm z_t)\), which approximates the clean latent variable associated with the current noisy state \(\bm z_t\). The clean latent estimate is decoded to a parameter field, passed through the surrogate forward model, evaluated at the observation locations, and compared with the data. The resulting data-misfit gradient provides the guidance direction for correcting the unconditional reverse step.

\begin{algorithm}[H]
\caption{Latent diffusion posterior-guided inversion}
\label{alg:ldps_compact}
\begin{algorithmic}[1]
\Require Decoder \(D_{\psi}\), diffusion model \(\bm\epsilon_{\omega}\), surrogate observation map \(\hat{\bm u}_o(\cdot)\), data \(\bm d\), noise level \(\gamma\), observation density \(\eta\), diffusion steps \(T\), guidance parameters \(\zeta_{\min},\zeta_{\max},c\)
\Ensure Approximate posterior-guided sample \(\hat{\kappa}\)
\State Sample \(\bm z_T \sim \mathcal{N}(0,I)\)
\For{\(t = T, T-1, \dots, 1\)}
    \State \(\hat{\bm\epsilon}_t \gets \bm\epsilon_{\omega}(\bm z_t,t)\)
    \State \(\hat{\bm z}_0 \gets \left(\bm z_t - \sqrt{1-\bar{\alpha}_t}\,\hat{\bm\epsilon}_t\right)/\sqrt{\bar{\alpha}_t}\)
    \State \(\hat{\kappa}_t \gets D_{\psi}(\hat{\bm z}_0)\)
    \State \(\hat{\bm u}_{o,t} \gets \hat{\bm u}_o(\hat{\kappa}_t)\)
    \State \(\mathcal L_{\mathrm{data}} \gets \|\bm d-\hat{\bm u}_{o,t}\|_2^2\)
    \State \(\bm g_t \gets \nabla_{\bm z_t}\mathcal L_{\mathrm{data}}\)
    \State Compute unconditional reverse step \(\bm z'_{t-1}\)
    \State \(\zeta \gets 
     \zeta_{\min}
        + (\zeta_{\max} - \zeta_{\min})\,
          \dfrac{\eta}{\eta + c\,\gamma^{2}}\)
    \State \(\bm z_{t-1} \gets \bm z'_{t-1} - \zeta\, \bm g_t/\|\bm g_t\|_2\)
\EndFor
\State \(\hat{\kappa} \gets D_{\psi}(\bm z_0)\)
\State \Return \(\hat{\kappa}\)
\end{algorithmic}
\end{algorithm}

Algorithm~\ref{alg:ldps_compact} shows that each reverse step requires two ingredients: a prior term and a likelihood-guidance term. The prior term is supplied by the diffusion model and keeps the evolving latent variable close to the learned latent distribution. The likelihood-guidance term is supplied by differentiating the data misfit and steers the sample toward consistency with the observations. We next derive these two terms.

For the noisy latent variable \(\bm z_t\), Bayes' rule gives
\begin{align}
    p_t(\bm z_t \mid \bm d)
    \propto
    p(\bm d \mid \bm z_t)\,p_t(\bm z_t),
    \label{eq:latent_bayes_noisy}
\end{align}
where \(p_t(\bm z_t)\) is the diffusion-time-dependent prior density of the noisy latent variable and \(p_t(\bm z_t\mid \bm d)\) is the corresponding posterior density conditioned on the data. The posterior score decomposes as
\begin{align}
    \nabla_{\bm z_t}\log p_t(\bm z_t\mid \bm d)
    =
    \nabla_{\bm z_t}\log p_t(\bm z_t)
    +
    \nabla_{\bm z_t}\log p(\bm d\mid \bm z_t).
    \label{eq:posterior_score}
\end{align}
The first term is the prior score, and the second term is the likelihood score.

The prior score is provided by the latent diffusion model trained in \cref{sec:ldps_prior}. Since the DDPM is trained to predict the Gaussian noise added in the forward diffusion process, the score of the noisy latent distribution is approximated by
\begin{align}
    \nabla_{\bm z_t}\log p_t(\bm z_t)
    \approx
    -
    \frac{1}{\sqrt{1-\bar{\alpha}_t}}\,
    \bm\epsilon_\omega(\bm z_t,t).
    \label{eq:latent_prior_score_dps}
\end{align}
Substituting \cref{eq:latent_prior_score_dps} into \cref{eq:posterior_score} gives
\begin{align}
    \nabla_{\bm z_t}\log p_t(\bm z_t\mid \bm d)
    \approx
    -
    \frac{1}{\sqrt{1-\bar{\alpha}_t}}\,
    \bm\epsilon_\omega(\bm z_t,t)
    +
    \nabla_{\bm z_t}\log p(\bm d\mid \bm z_t).
    \label{eq:latent_dps_score}
\end{align}
Thus, posterior-guided sampling requires an approximation of the likelihood score
\(\nabla_{\bm z_t}\log p(\bm d\mid \bm z_t)\).

This likelihood score is not available directly because the observations are related to the clean parameter field, or equivalently to the clean latent variable \(\bm z_0\), whereas the reverse diffusion process evolves the noisy variable \(\bm z_t\). Following DPS~\cite{chung2023diffusion}, we use the Tweedie denoised estimate
\begin{align}
    \hat{\bm z}_0(\bm z_t)
    =
    \frac{
        \bm z_t
        -
        \sqrt{1-\bar{\alpha}_t}\,
        \bm\epsilon_\omega(\bm z_t,t)
    }{
        \sqrt{\bar{\alpha}_t}
    },
    \label{eq:tweedie_latent}
\end{align}
and adopt the plug-in approximation
\begin{align}
    p(\bm d\mid \bm z_t)
    \approx
    p(\bm d\mid \hat{\bm z}_0(\bm z_t)).
    \label{eq:plug_approx}
\end{align}
This approximation converts the intractable likelihood of the noisy latent variable into a tractable likelihood evaluated at an estimated clean latent variable.

In the proposed latent PDE setting, the clean latent estimate is first decoded to a parameter field,
\begin{align}
    \hat{\kappa}_t
    =
    D_\psi\!\left(\hat{\bm z}_0(\bm z_t)\right).
\end{align}
The corresponding surrogate-predicted observations are then
\begin{align}
    \hat{\bm u}_{o,t}
    =
    \hat{\bm u}_o(\hat{\kappa}_t)
    =
    \hat{\bm u}_o
    \left(
        D_\psi[\hat{\bm z}_0(\bm z_t)]
    \right),
    \label{eq:predicted_observations_latent}
\end{align}
where \(\hat{\bm u}_o(\kappa)\) denotes the neural-surrogate prediction of the PDE state at the observation locations for a given parameter field \(\kappa\).

Under the Gaussian observation model in \cref{eq:obs}, the plug-in likelihood is
\begin{align}
    p(\bm d\mid \hat{\bm z}_0(\bm z_t))
    \propto
    \exp\left[
        -
        \frac{1}{2\sigma^2}
        \left\|
            \bm d-\hat{\bm u}_{o,t}
        \right\|_2^2
    \right].
\end{align}
Therefore,
\begin{align}
    \nabla_{\bm z_t}\log p(\bm d\mid \bm z_t)
    \approx
    -
    \frac{1}{2\sigma^2}
    \nabla_{\bm z_t}
    \left\|
        \bm d-\hat{\bm u}_{o,t}
    \right\|_2^2.
    \label{eq:likelihood_score_latent}
\end{align}
The gradient in \cref{eq:likelihood_score_latent} is obtained by differentiating the data misfit through the chain
\begin{align}\label{eq:chain}
    \bm z_t
    \rightarrow
    \hat{\bm z}_0(\bm z_t)
    \rightarrow
    D_\psi(\hat{\bm z}_0)
    \rightarrow
    \hat{\bm u}_o
    \left(
        D_\psi[\hat{\bm z}_0(\bm z_t)]
    \right).
\end{align}
This is the main difference between standard DPS and the proposed L-DPS formulation: standard DPS evaluates the likelihood via the measurement operator in the ambient unknown space, whereas L-DPS evaluates it via the decoder and surrogate-predicted observations. The diffusion model provides the latent prior score, while the differentiable surrogate provides likelihood guidance without repeated PDE solves.

For use in the stabilized update, we define the data-misfit gradient
\begin{align}
    \bm g_t(\bm z_t)
    =
    \nabla_{\bm z_t}
    \left\|
        \bm d-\hat{\bm u}_{o,t}
    \right\|_2^2.
    \label{eq:data_misfit_gradient}
\end{align}
The approximate Gaussian likelihood score is proportional to
\(-\bm g_t/(2\sigma^2)\). In the implementation below, we use the direction of \(\bm g_t\) to guide the reverse-diffusion update and to control the step length separately via a stabilized guidance weight.

The likelihood correction in \cref{eq:likelihood_score_latent} should not be interpreted as an exact posterior transition. It relies on the learned latent representation, the DPS plug-in approximation in \cref{eq:plug_approx}, the surrogate approximation of the forward model, and a finite-step reverse diffusion discretization. The proposed method is therefore an approximate posterior-guided inversion procedure rather than an exact posterior sampler.

Algorithm~\ref{alg:ldps_compact} produces one approximate posterior sample of the parameter field. 
To generate an ensemble as might be required for uncertainty quantification, the steps 1-14 of the algorithm must be repeated \(S\) times, using independent initial latent variables
\[
    \bm z_T^{(s)} \sim \mathcal N(0,I), 
    \qquad s=1,\ldots,S,
\]
and the same observations \(\bm d\), surrogate observation map \(\hat{\bm u}_o(\cdot)\), and guidance parameters. 
Each reverse trajectory produces a terminal latent sample \(\bm z_0^{(s)}\), which is decoded to obtain
\[
    \kappa^{(s)} = D_\psi(\bm z_0^{(s)}),
    \qquad s=1,\ldots,S.
\]
The collection
\[
    \left\{\kap^{(s)}\right\}_{s=1}^S
\]
is interpreted as an approximate ensemble from the posterior distribution \(p(\kap\mid \bm d)\), subject to the approximations described above.

In the numerical experiments reported below, however, we set \(S=1\) and report a single posterior-guided reconstruction for each test case. 
Thus, the present work evaluates reconstruction accuracy and computational efficiency rather than posterior uncertainty quantification.

\subsection{Stabilized Guidance}
\label{sec:stabilized_guidance}

The data-misfit gradient in \cref{eq:data_misfit_gradient} provides the direction in latent space that most rapidly reduces the discrepancy between the observed data and the surrogate-predicted observations. However, using the raw gradient magnitude directly can destabilize the reverse diffusion process. The magnitude of \(\bm g_t\) varies across diffusion steps because it depends on the observation residual, the diffusion noise level, and the Jacobians of both the decoder and the surrogate observation map. Consequently, the raw magnitude of \(\bm g_t\) is not a reliable measure of how far the latent variable should be moved at a given reverse diffusion step.

To separate the direction of the likelihood correction from its step length, we use the normalized guidance update
\begin{align}
    \bm z_{t-1}
    =
    \bm z'_{t-1}
    -
    \zeta\,
    \frac{\bm g_t}{\|\bm g_t\|_2},
    \label{eq:stabilized_update}
\end{align}
where \(\bm z'_{t-1}\) is the latent variable obtained from the unconditional reverse diffusion step and \(\zeta\) is the guidance step size. The unconditional step incorporates the learned prior through the diffusion model, while the second term in \cref{eq:stabilized_update} moves the sample in the direction that improves data consistency. By normalizing the guidance direction, the magnitude of the update is controlled explicitly by \(\zeta\), rather than implicitly by the possibly ill-scaled decoder--surrogate gradient. A related normalization is used in the original DPS algorithm~\cite{chung2023diffusion}; here, we use gradient-magnitude normalization because it is more robust to the heterogeneous scaling introduced by the decoder and surrogate.

The guidance strength should depend on both the noise level and the amount of observational information. If the data are noisy, strong guidance can overfit the observational noise. If the observations are sparse, the likelihood contains limited information relative to the dimension of the unknown field, and the update should rely more strongly on the learned prior. A formal Gaussian likelihood suggests that guidance strength is proportional to the inverse of the noise variance. Under the observation model in \cref{eq:obs}, this scaling is proportional to \(N_{\mathrm{obs}}/\gamma^2\). In practice, this scaling is too aggressive for L-DPS. First, the likelihood is evaluated at the Tweedie estimate \(\hat{\bm z}_0(\bm z_t)\), which is inaccurate at early reverse steps. Scaling the guidance by \(1/\gamma^2\) can therefore amplify errors in the plug-in approximation. Second, the factor \(N_{\mathrm{obs}}/\gamma^2\) can become excessively large for small \(\gamma\) or large \(N_{\mathrm{obs}}\), pushing the sample away from the learned latent manifold. This can occur because the guidance direction is already normalized in \cref{eq:stabilized_update}. 

We therefore use a bounded, noise- and density-adaptive guidance weight. Let
\begin{align}
    \eta = \frac{N_{\mathrm{obs}}}{N_{\mathrm{tot}}},
\end{align}
where \(N_{\mathrm{tot}}\) is the total number of admissible observation locations. Validation experiments show two limiting regimes. In the clean-and-dense regime, the optimal guidance weight saturates at a maximum value because the observations provide reliable information and can be enforced strongly. In the noisy-or-sparse regime, the optimal guidance weight approaches a small positive floor, as the data still contains useful information but should not dominate the learned prior. We interpolate between these regimes using
\begin{align}
    \zeta(\eta,\gamma)
    =
    \zeta_{\min}
    +
    (\zeta_{\max}-\zeta_{\min})
    \frac{\eta}{\eta+c\gamma^2},
    \label{eq:adaptive_zeta}
\end{align}
where \(\zeta_{\max}\) is the clean-and-dense cap, \(\zeta_{\min}\) is the noisy-or-sparse floor, and \(c\) controls the transition between the two regimes. This function is monotone increasing in the observation density \(\eta\), monotone decreasing in the relative noise level \(\gamma\), and remains bounded for both \(\gamma\to 0\) and large \(N_{\mathrm{obs}}\). The parameters \((\zeta_{\max},\zeta_{\min},c)\) are calibrated once on the validation set, as described in \cref{sec:ablation_guidance}.

The resulting stabilized update combines three components: the unconditional reverse diffusion step, which represents the learned latent prior; the normalized data-misfit gradient, which supplies the likelihood direction; and the adaptive scalar weight \(\zeta(\eta,\gamma)\), which determines how strongly the observations influence each reverse step.

\subsection{Interpretation}
\label{sec:interpretation_algorithm}

The proposed method should be interpreted as an approximate posterior-guided inversion framework rather than an exact posterior sampler.  The reverse diffusion model supplies an approximation to the prior score in latent space. The decoder maps the clean latent estimate to a parameter field. The surrogate observation map \(\hat{\bm u}_o(\cdot)\) maps this parameter field to predicted observations. The data mismatch produces a likelihood-guidance direction. Finally, the stabilized update combines the prior-driven reverse diffusion step with the normalized likelihood guidance.

The method introduces four approximations for computational tractability:
\begin{enumerate}[nosep]
    \item \textbf{Representation restriction}: the parameter field is restricted to the range of the VAE decoder \(D_\psi\).
    \item \textbf{Plug-in likelihood}: the intractable likelihood \(p(\bm d\mid \bm z_t)\) is approximated by evaluating the likelihood at the Tweedie estimate \(\hat{\bm z}_0(\bm z_t)\).
    \item \textbf{Surrogate replacement}: the full PDE forward model is replaced by the differentiable surrogate observation map \(\hat{\bm u}_o(\cdot)\), so that likelihood gradients can be computed without repeated numerical PDE solves.
    \item \textbf{Finite-step stabilized guidance}: the posterior correction is applied through a discrete reverse diffusion process using normalized guidance directions and an adaptive step size.
\end{enumerate}

The offline and online stages are separated. Offline, we train the VAE to obtain a compact latent representation of the parameter field, train the latent diffusion model to represent the prior score of the encoded variables, and train the neural surrogate to approximate the PDE solution at the observation locations. Online, given observations \(\bm d\), we initialize \(\bm z_T\sim\mathcal N(0,I)\) and run the guided reverse diffusion process in Algorithm~\ref{alg:ldps_compact}. At each step, the prior score keeps the sample on the learned latent manifold, while the stabilized guidance term steers the sample toward consistency with the observations. The final estimate is obtained by decoding the terminal latent sample,
\begin{align}
    \hat{\kappa}=D_\psi(\bm z_0).
\end{align}

Compared with standard DPS, which operates directly in the original unknown space and evaluates the measurement operator there, the proposed L-DPS method performs the reverse diffusion and guidance in a reduced latent space. It evaluates the likelihood through the decoded parameter field and surrogate-predicted observations. This combination makes DPS practical for high-dimensional PDE inverse problems with implicit non-Gaussian priors and expensive forward models.

\begin{figure}[htbp]
    \centering
    \includegraphics[width=0.95\textwidth]{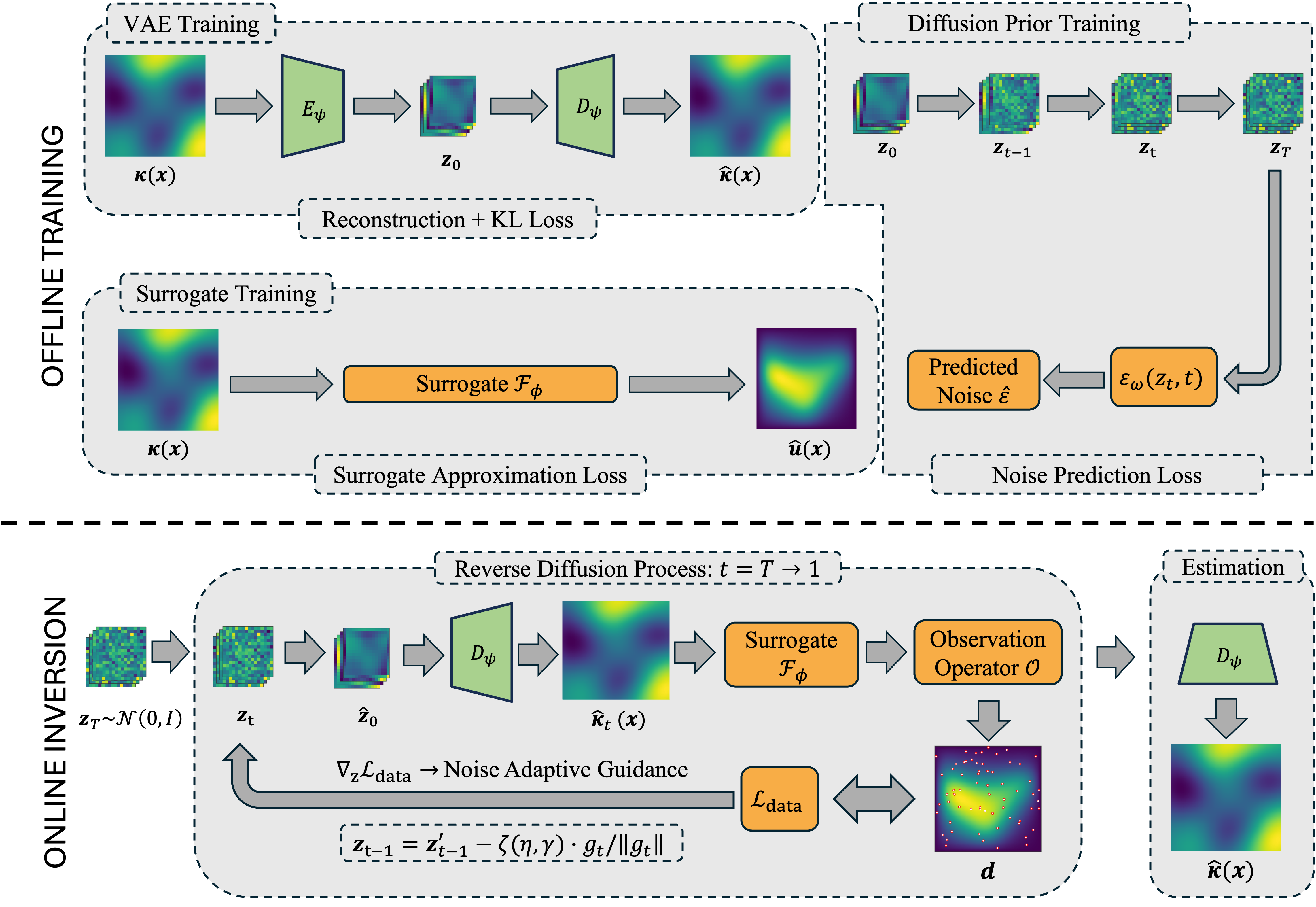}
\caption{Overview of the L-DPS framework. The VAE is trained to reconstruct the parameter field $\kappa(x)$ with the encoder $E_\psi$ and decoder $D_\psi$; the latent diffusion prior $\epsilon_\omega(z_t,t)$ is trained on the encoded latents $z_0$; and the neural surrogate $\mathcal{F}_\phi$ (FNO / DeepONet / ViT) is trained to approximate the forward PDE solver $\mathcal{F}$ on the full pressure field $\hat u(x)$. At inference, the reverse diffusion process (bottom) starts from $z_T$, repeatedly computes the Tweedie estimate $\hat z_0$, decodes to $\hat\kappa(x)$, evaluates the data-misfit $\mathcal{L}_{\mathrm{data}}$ against the sparse observations $d$, and applies the noise-adaptive guidance $z_{t-1}=z'_{t-1}-\zeta(\eta,\gamma)\,g_t/\|g_t\|$ until $t=0$; the final decoded field $\hat\kappa(x)$ is the estimate.}
    \label{fig:ldps_overview}
\end{figure}

\section{Case Study: Steady State Darcy problem}\label{sec:darcy_case}
\subsection{Problem setting}\label{sec:darcy_problem}
We assess the proposed framework on the inverse Darcy flow problem with an unknown space-dependent permeability field. The Darcy flow problem is commonly used to test methods for high-dimensional inverse PDEs.

Let $\Omega \subset \mathbb{R}^2$ be the computational domain. The pressure field $u(\bm{x})$ satisfies the steady-state Darcy equation
\begin{align}
    -\nabla \cdot \bigl(\kappa(\bm{x}) \nabla u(\bm{x})\bigr) = f_s(\bm{x}), \qquad \bm{x}\in\Omega, \label{eq:darcy}
\end{align}

subject to appropriate boundary conditions. Here, $u(\bm{x})$ is the hydraulic head, $\kappa(\bm{x})$ is the conductivity field, and $f_s(\bm{x})$ is the source term. In the inverse problem, we set $\Omega=[0,1]^2$, $f_s(\bm{x})=1$, zero Dirichlet boundary conditions on all boundaries, and estimate $\kap$, the discretized parameter field, from sparse noisy observations of $u(\bm{x})$ denoted as $\bm{d}$. 
%
This inverse problem is challenging because $\kap$ is high-dimensional, whereas $\bm{d}$ is low-dimensional. In addition, repeated evaluations of the forward Darcy solution in posterior-guided inference are relatively expensive. These features make the inverse Darcy flow problem a suitable benchmark for assessing the proposed computational framework.

\subsection{Dataset generation, observation settings, and reporting of the results}

We generate 5000 log-Gaussian fields $\bm{y}=\log\kap$ with correlation length $l=0.2$ and $\sigma_{\bm{y}}=1.0$ on a $128\times128$ uniform grid, split 4000/500/500 for training/validation/testing. The Darcy equation \cref{eq:darcy} is solved with a finite-difference scheme to produce paired $(\kap, u)$ data.

Sensors are placed at random interior grid points (using a fixed seed for reproducibility). The observation noise in \cref{eq:obs} is set to
\begin{align}
  \boldsymbol{\epsilon} \;\sim\; \mathcal{N}\!\left(0,\,
    \gamma^2 \frac{\| \bm{d}\|_2^{2}}{N_{\mathrm{obs}}}\,\mathbf{I}\right),
  \label{eq:noise_model}
\end{align}
so that $\gamma$ is the relative noise level. In the numerical experiments we vary $\gamma\in\{0,\,0.01,\,0.02,\,0.05,\,0.1\}$ and $N_{\mathrm{obs}}\in\{16,\,32,\,64,\,128,\,256,\,512\}$.

We report the accuracy of the inverse solution in terms of the relative $\ell_2$ error in the estimated $\bm{y}$ as,
\begin{align}
  \ell_2 \;=\; \frac{\|\hat{\bm{y}} - \bm{y}_r\|_2}{\|\bm{y}_r\|_2},\label{eq:rl2_metric}
\end{align}
averaged over 500 test samples. Here, $\bm{y}_r$ and $\hat{\bm{y}}$ are the reference and estimated parameter vectors, respectively.  We also compare the estimated and reference fields.  No ensemble-averaged fields are shown unless stated otherwise.

\subsection{Model architectures and training}

In the latent DPS inverse solution and other considered inverse methods, we approximate the forward map from $\kap$ to $u(\bm{x})$ and, ultimately, $\bm{u}_o$, the PDE solution vector at the measurement locations, using operator learning surrogate models. Three surrogates are evaluated: FNO (4.2M parameters, 0.46\% test relative $L_2$), ViT (6.0M parameters, 0.23\%), and DeepONet (1.3M parameters, 0.72\%). The DDPM of the latent prior is based on a VAE with $\beta=10^{-6}$ (4.0M parameters, 0.3\% reconstruction error) and a latent UNet model (5.3M parameters). All ML models (including surrogate models, VAE, and the latent-UNet DDPM model) are trained with a cosine learning-rate schedule, early-stopping patience of 200-250, and up to 5000 epochs.

Posterior-guided inversion uses 1000 DDIM~\cite{song2020denoising} reverse-diffusion steps. The noise- and density-adaptive guidance weight is given by \cref{eq:adaptive_zeta} with $\zeta_{\max} = 0.5$, $\zeta_{\min} = 0.04$, and $c = 38$. These values are obtained from the validation-set grid search described in \cref{sec:ablation_guidance} and are used throughout the numerical experiments unless noted otherwise. For each test case, we generate one sample of the parameter field from its posterior, i.e., we set $S=1$ in the L-DPS algorithm. Uncertainty quantification (UQ) of the inverse solution is not the focus of this paper. However, a UQ study can be performed straightforwardly by setting $S >> 1$.

\subsection{Baseline methods}

We compare the proposed L-DPS framework with four methods, each representing a different family of PDE-inversion methods reported in the literature: full-$\kap$-space DPS, direct inverse operators, conditional latent-space diffusion model, and Karhunen--Lo\`eve expansion (KLE)-defined latent-space MAP (KLE-MAP) method. All these methods use the same dataset and, where applicable, the same forward FNO surrogate so that the comparison isolates the effect of the inversion strategy.

\begin{itemize}[nosep]
\item \textbf{Full-space DPS + FNO}: A DDPM model (with $17.5$M parameters) of $\kap$ prior distribution is trained directly on the samples $\kap^{(i)} \in \mathbb{R}^{1 \times 128\times128}$. Posterior sampling uses the same surrogate-based likelihood guidance as L-DPS, including the same adaptive weight \cref{eq:adaptive_zeta} and the same stabilized update \cref{eq:stabilized_update}; the only difference from L-DPS is that the reverse diffusion and the guidance act on $\kap_t$ instead of on the latent $z_t$. The UNet in full-space DDPM follows the same training protocol as the latent-space DDPM (cosine schedule, early-stopping patience $250$, up to $5000$ epochs). This baseline isolates the effect of the latent representation.
\item \textbf{Direct Inverse FNO (InvFNO)}: A direct regression network (with $4.2$M trainable parameters) that maps the sparse pressure observations $\bd \in \mathbb{R}^{N_{\mathrm{obs}}}$ to the log-permeability field $\log\kappa$. A new network must be trained whenever the location and number of measurements change. 
\item \textbf{Conditional Latent-Space Diffusion Model (CLDM)}: A conditional latent diffusion model trained on (observation, parameter) pairs for one specific $N_{\mathrm{obs}}$, which learns the posterior by amortization rather than by posterior guidance. A separate model is trained for each $N_{\mathrm{obs}}$, so the method does not generalize across sensor layouts. This baseline represents the amortized-posterior family of diffusion-based inverse methods.
\item \textbf{KLE-MAP}: A parameter estimation method that uses Karhunen--Lo\`eve expansion (KLE) to define the latent parameter space and uses the maximum a posteriori (MAP) estimate of $\kap$ in this latent space. The (forward) FNO model is used to compute the log likelihood in the MAP objective function. 
For Gaussian random fields, the KLE provides the optimal linear mean-square representation, with Gaussian-distributed latent coefficients.
Therefore,  this method serves as the baseline for L-DPS for inverse problems with Gaussian parameter priors. Compared in \cref{sec:ablation_kle}.
\end{itemize}

In addition to these four baselines, we consider five alternative guidance strategies applied to the same L-DPS + FNO pipeline: a constant $\zeta = 0.5$, the original DPS guidance of \cite{chung2023diffusion}, a Bayesian scaling in which $\zeta$ is set to the inverse of the noise variance, JAPS, and DiffStateGrad. These guidance variants share the same prior and surrogate with L-DPS and differ only in the guidance rule; they are reported in \cref{sec:ablation_guidance}. To assess how surrogate accuracy affects inversion performance, we compute L-DPS reconstructions using two additional neural-operator surrogates, DeepONet and ViT, alongside the FNO surrogate.

\subsection{Comparison of L-DPS with benchmark inverse methods}\label{sec:main_results}

\begin{figure}[htbp]
\centering
\includegraphics[width=\textwidth]{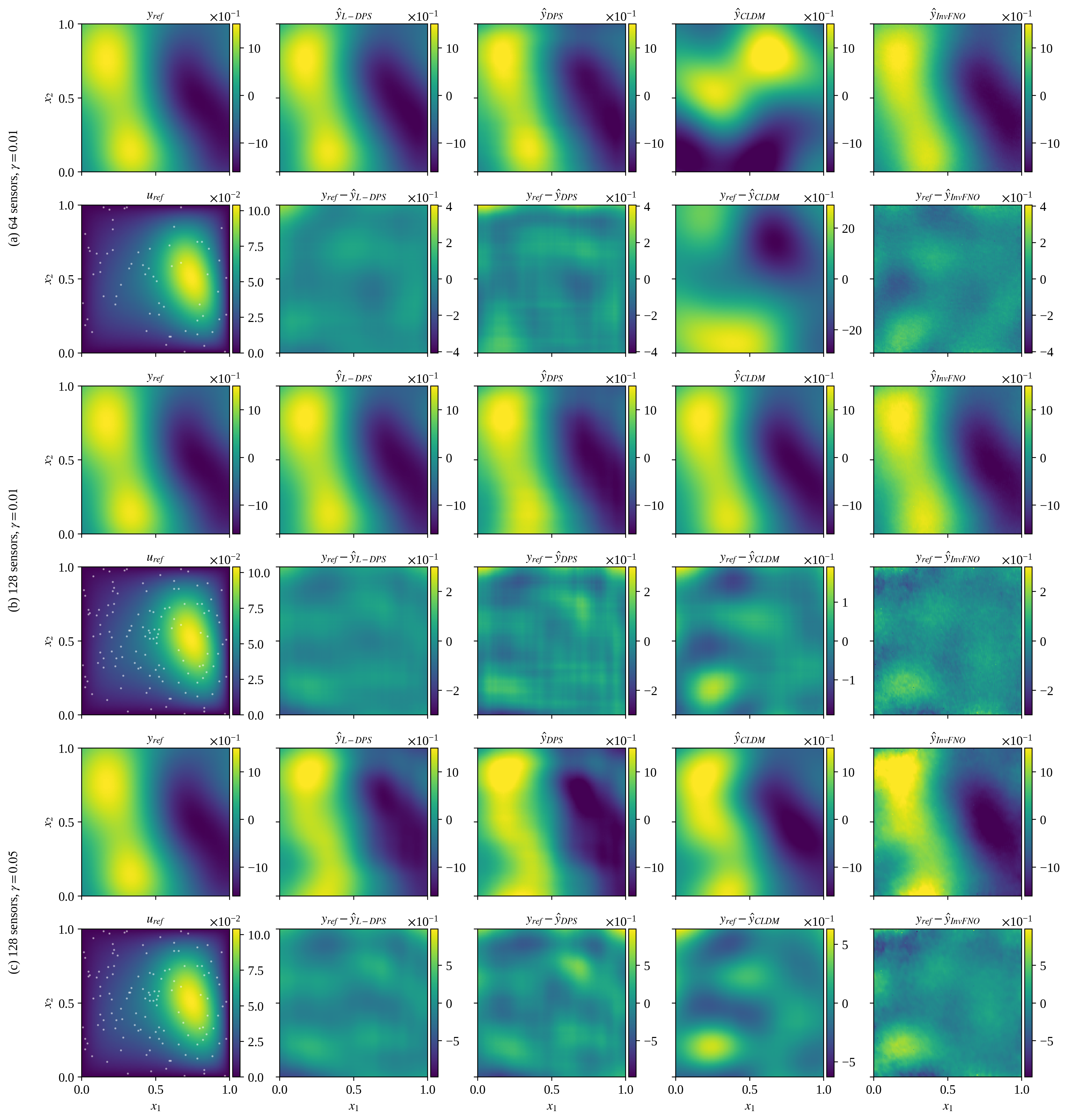}
\caption{Comparison of the reference parameter field ($\bm{y}_{ref}$) and the fields estimated with the L-DPS+FNO ($\hat{\bm{y}}_{L-DPS}$), full-space DPS ($\hat{\bm{y}}_{DPS}$), CLDM  ($\hat{\bm{y}}_{CLDM}$), and InvFNO ($\hat{\bm{y}}_{InvFNO}$) methods. Also shown are the reference $\bm{u}$ solution ($\bm{u}_{ref}$) and the point errors between estimated $\bm{y}$ and $\bm{y}_{ref}$.  The inverse solutions are obtained under three settings (different rows): (a) The number of observations $N_{obs}=64$ and the relative noise level $\gamma=0.1$, (b) $N_{obs}=128$ and $\gamma=0.01$, and (c) $N_{obs}=128$ and $\gamma=0.05$. 
CLDM point errors are plotted with a different color scale than the other three methods because they have a significantly smaller range. 
}
\label{fig:qualitative}
\end{figure}

Figure~\ref{fig:qualitative} provides a qualitative comparison of the inverse reconstructions obtained with L-DPS+FNO and the three benchmark methods: full-space DPS, CLDM, and InvFNO. For each test case, the figure shows the reference log-conductivity field \(\bm y_{\mathrm{ref}}\), the estimated fields, and the corresponding pointwise reconstruction errors with respect to \(\bm y_{\mathrm{ref}}\). The PDE solution \(\bm{u}_{\mathrm{ref}}\) and the observation locations are also shown in the reference column. The three rows correspond to increasingly different observation regimes: (a) sparse and noisy observations, \(N_{\mathrm{obs}}=64\) and \(\gamma=0.1\); (b) the base setting, \(N_{\mathrm{obs}}=128\) and \(\gamma=0.01\); and (c) moderate noise, \(N_{\mathrm{obs}}=128\) and \(\gamma=0.05\). The CLDM errors are plotted using a separate color scale because their error range differs substantially from those of the other methods.

The qualitative results show that L-DPS+FNO provides the most stable reconstructions across the three regimes. In the sparse-and-noisy case, L-DPS+FNO recovers the dominant spatial structure of the reference field. In contrast, full-space DPS and InvFNO exhibit larger structured errors, and CLDM produces a less reliable reconstruction. In the base setting, all methods recover the main features of the field, but L-DPS+FNO gives the sharpest reconstruction and the smallest visual error. Under moderate noise, the benefit of the latent prior and stabilized likelihood guidance becomes more apparent: L-DPS+FNO preserves the large-scale field structure while avoiding the stronger artifacts and smoothing observed in the benchmark reconstructions.

\FloatBarrier

\begin{figure}[htbp]
\centering
\includegraphics[width=\textwidth]{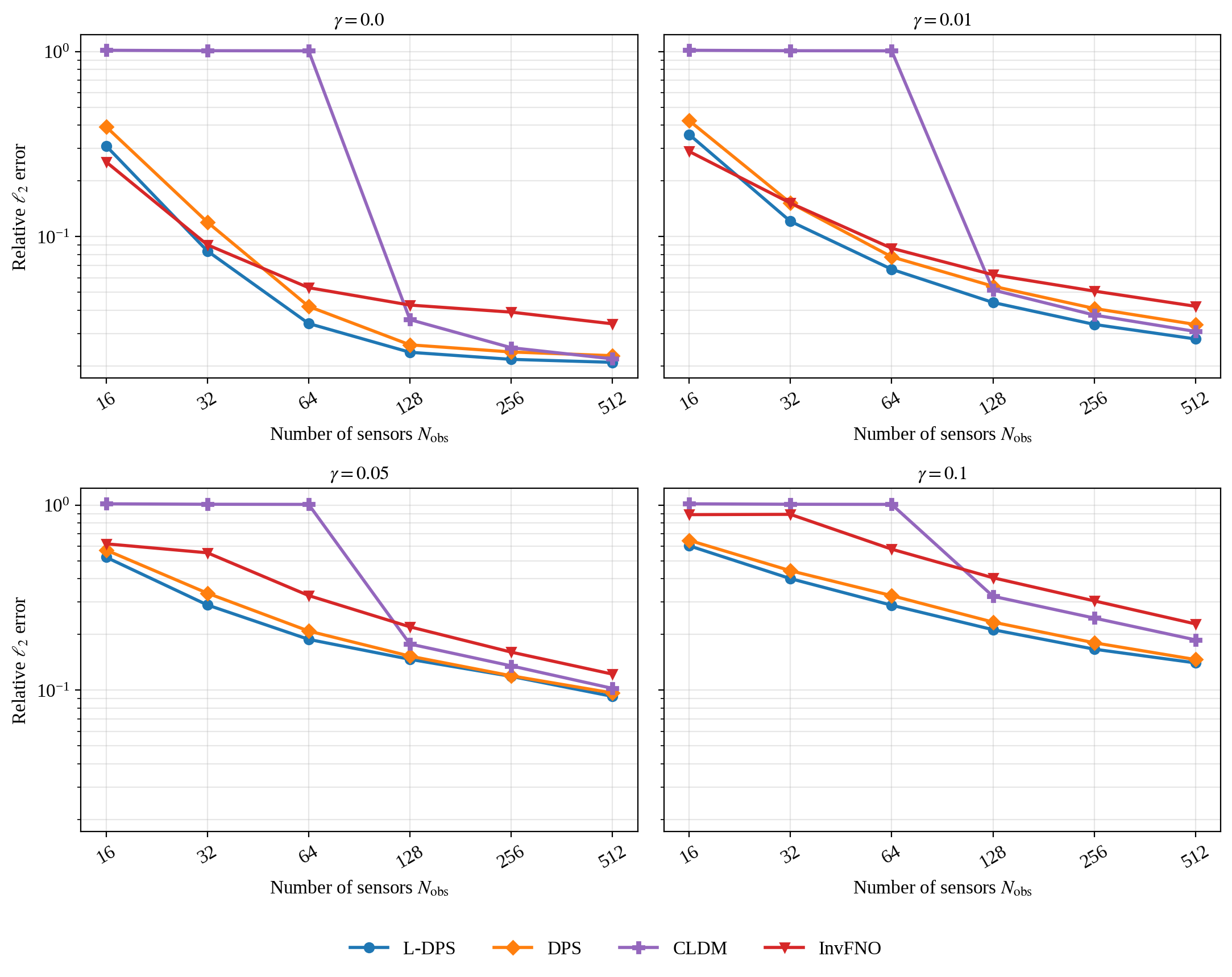}
\caption{Relative $\ell_2$ reconstruction error on $y=\log\kappa$ versus number of sensors $N_{\mathrm{obs}}$ across four noise levels $\gamma\in\{0,0.01,0.05,0.1\}$ (one panel each). L-DPS+FNO consistently matches or beats the baselines across the grid; the gap widens in the sparse/noisy regime, where amortized methods (CLDM, InvFNO) and full-space DPS degrade noticeably.}
\label{fig:error_vs_sensors}
\end{figure}

Figure~\ref{fig:error_vs_sensors} summarizes the quantitative performance by plotting the mean relative \(\ell_2\) error in the estimated \(\bm y\) field as a function of the number of observations \(N_{\mathrm{obs}}\) for four noise levels, \(\gamma\in\{0,0.01,0.05,0.1\}\). Across most observation and noise settings, L-DPS+FNO matches or outperforms the benchmark methods. The advantage is especially pronounced in noisy regimes, where enforcing data consistency without a sufficiently informative prior can lead to overfitting to noisy observations. For example, at \(N_{\mathrm{obs}}=256\) and \(\gamma=0.1\), L-DPS+FNO obtains a relative \(\ell_2\) error of \(0.178\), compared with \(0.446\) for full-space DPS and \(0.304\) for InvFNO. In contrast, under noiseless observations, full-space DPS is competitive with L-DPS+FNO, indicating that strong data consistency can compensate for the weaker full-space prior when measurements are error-free. Overall, the results in Figures~\ref{fig:qualitative} and~\ref{fig:error_vs_sensors} indicate that the latent representation and the stabilized surrogate-based guidance are most beneficial when observations are sparse and/or noisy.

\FloatBarrier


\subsection{Effect of the latent representation}\label{sec:ablation_latent}

To isolate the effect of the latent representation, we first focus on the two DPS-based methods from the benchmark comparison in Section~\ref{sec:main_results}: L-DPS+FNO and full-space DPS+FNO. We compare the performance of the two models for measurements with the different noise levels $\gamma$.
Both methods use the same FNO surrogate, adaptive guidance rule, observation model, and diffusion training protocol; they differ only in the space in which the diffusion prior and posterior-guided sampling are defined. 

For the Darcy problem, the unknown log-conductivity field is a scalar field on a \(128\times128\) grid and is represented as a single-channel tensor,
\[
    \bm y \in \mathbb{R}^{C_\kappa\times H\times W}
    =
    \mathbb{R}^{1\times128\times128},
\]
with \(C_\kappa=1\) and \(1\times128\times128=16{,}384\) degrees of freedom. 
In the VAE encoder, we set \(f=8\) and \(C_z=4\), so that $\bm{y}$ is mapped to a lower-resolution latent feature tensor,
\[
    \bm z \in \mathbb{R}^{C_z\times H/f\times W/f}
    =
    \mathbb{R}^{4\times16\times16}.
\]

Table \ref{tab:latent_vs_pixel} reports the inverse relative $\ell_2$ error at $N_{\mathrm{obs}} = 256$ for the two methods. Under noiseless observations, the two methods achieve comparable accuracy: L-DPS gives $0.022$, and full-space DPS gives $0.023$. When the observations are noiseless and dense, the surrogate-based guidance is strong enough to pull either representation close to the truth, and the choice of the representation space is secondary. The gap between the two methods widens once noise is added. At $N_{\mathrm{obs}} = 256$ and $\gamma = 0.05$, full-space DPS achieves $0.148$, while L-DPS achieves $0.127$. At $\gamma = 0.1$, full-space DPS degrades to $0.229$, while L-DPS reaches $0.178$, a $22\%$ relative improvement. The same pattern holds at fixed $\gamma = 0.02$ across sensor counts: at $N_{\mathrm{obs}} = 64$, full-space DPS produces a $43\%$ larger error than L-DPS ($0.174$ vs.\ $0.122$).

Three observations follow from Table \ref{tab:latent_vs_pixel} and the trends in Figure \ref{fig:error_vs_sensors}. First, compressing the sampling space by a factor of 16 (16{,}384 $\to$ 1024) does not sacrifice accuracy when using noiseless data--the latent representation is rich enough to encode the smooth log-permeability fields used here. Second, the latent prior regularizes more effectively under observation noise: because the VAE concentrates prior mass on a smooth manifold, a noisy likelihood gradient that would otherwise drag the full-space sample off the plausible set is projected back onto the learned manifold in the latent space. Third, operating in 1024 dimensions rather than 16384 reduces the per-step computational cost by roughly a factor of five, yielding a combined benefit in accuracy and efficiency.

Finally, we study the accuracy of the VAE latent-space representation to determine whether the VAE encoder--decoder pair is the limiting factor in inverse solution accuracy. Specifically, we study the VAE's accuracy in reconstructing $\kap$ from its latent representation $\bm{z}_0$.  For this test, we consider three $\bm{y}$ fields with three different prior distributions: the Gaussian with a single-correlation length exponential covariance function (baseline with the correlation length $l=0.2$ considered in Section \ref{sec:main_results}), a multiscale Gaussian (a superposition of two Gaussian random fields with the exponential covariance functions with correlation lengths $l=0.4$ and $l=0.2$), and a binary random field generated by transforming the Gaussian $\tilde{\bm{y}}$  with zero mean, Matérn covariance function, and the correlation length $l=0.2$ as $y_{ij}=2.485$ ($\kappa_{ij}=12$) for $\tilde{y}_{ij} \ge 0$ and $y_{ij}=1.386$ ($\kappa_{ij} = 4$) for $\tilde{y}_{ij} < 0$.

\begin{figure}[htbp]
\centering
\includegraphics[width=0.95\textwidth]{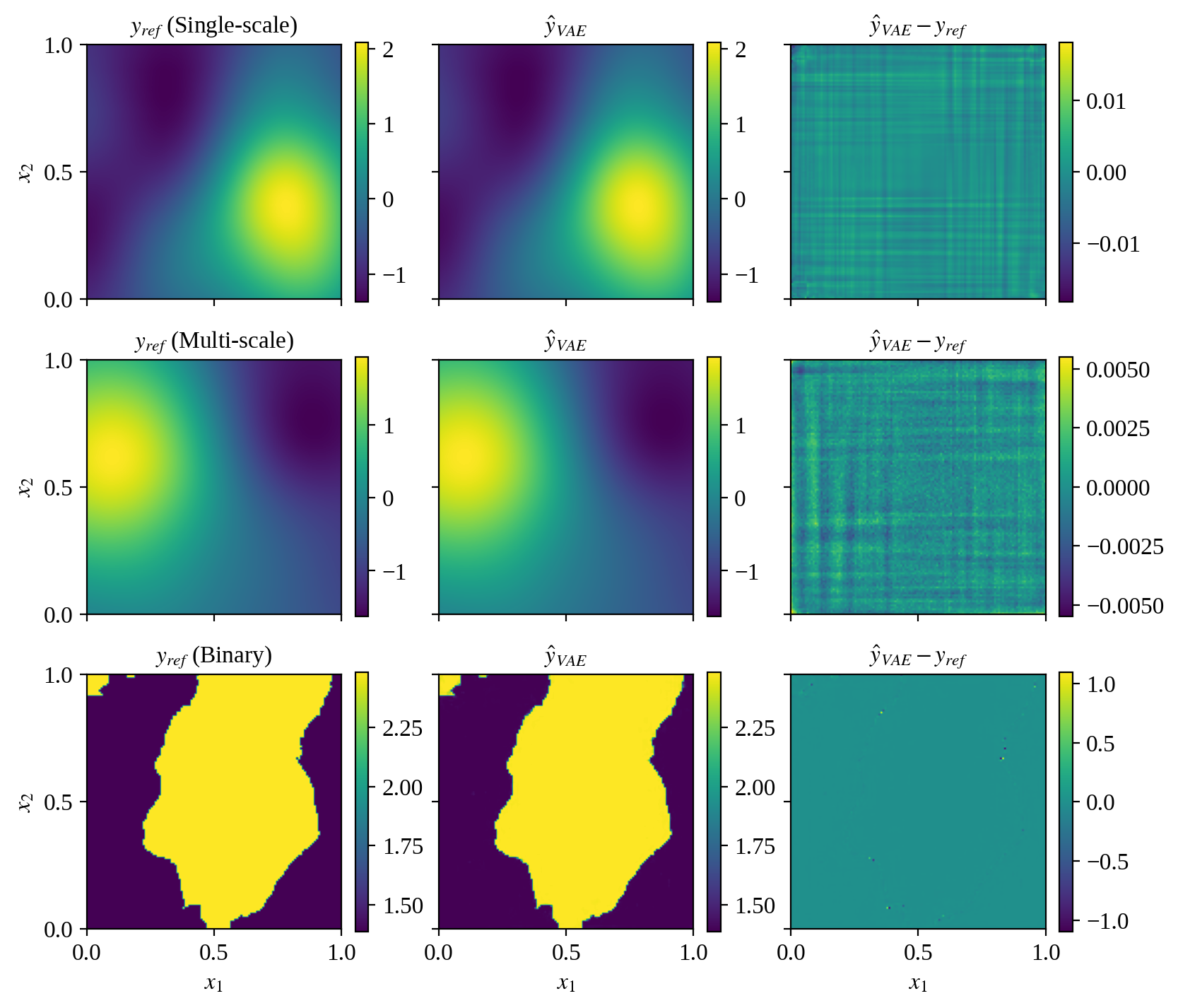}
\caption{VAE reconstruction quality for three priors. Rows: single-scale Gaussian, multi-scale Gaussian, and binary non-Gaussian. Columns: reference $\kappa$, VAE reconstruction, and the reconstruction error. Each row shows the median-error test sample.}
\label{fig:vae_recon_three}
\end{figure}

\begin{figure}[htbp]
\centering
\includegraphics[width=0.7\textwidth]{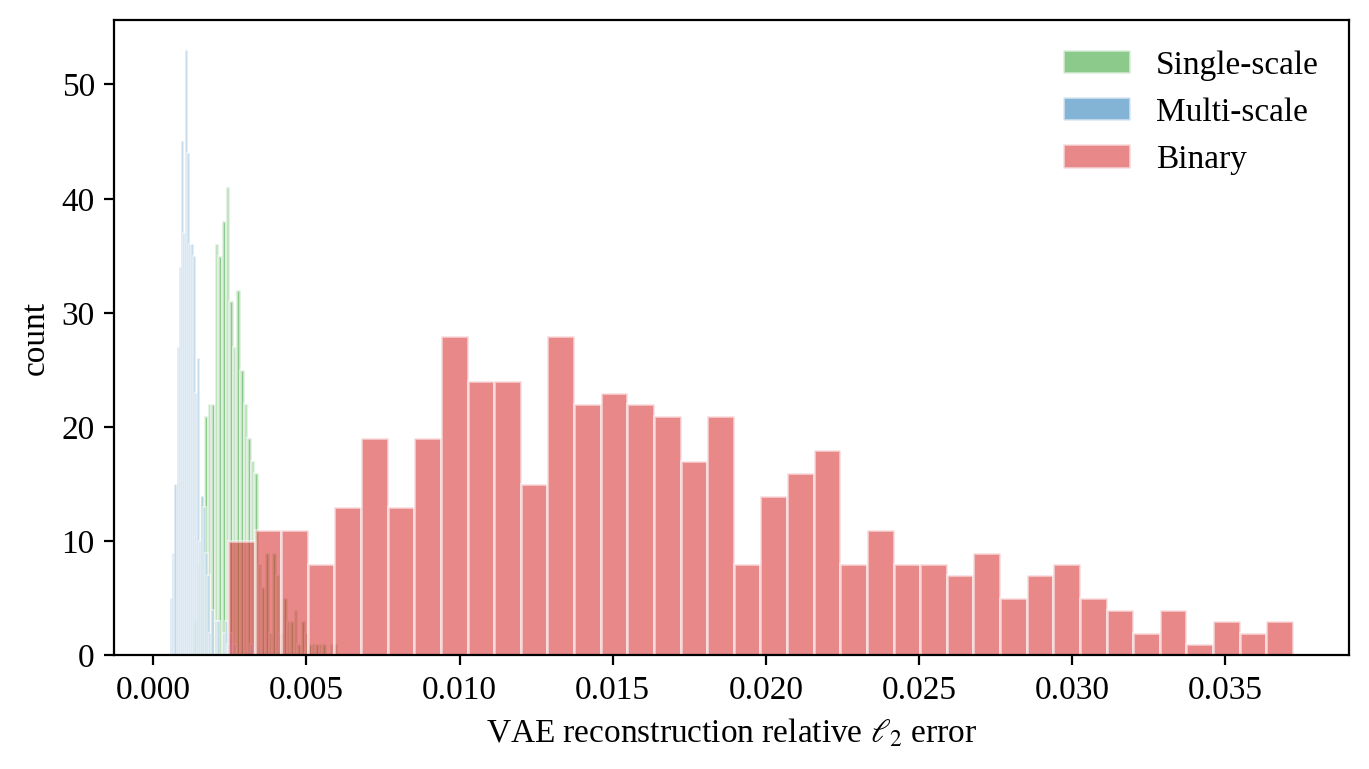}
\caption{Distribution of the VAE reconstruction relative $\ell_2$ error over $500$ test samples for each prior. The three distributions are concentrated well below $4\%$.}
\label{fig:vae_recon_hist}
\end{figure}

We generate 5000 samples from each distribution (4000 for training, 500 for validation, and 500 for testing) and train three VAE models, one for each sample set. We perform reconstruction and compute the $\ell_2$ reconstruction error for each sample in the test set.     
Figure \ref{fig:vae_recon_three} shows the reconstructed $\kap$ corresponding to the median $\ell_2$ reconstruction error for the three priors considered in this work. 
Figure \ref{fig:vae_recon_hist} shows the distribution of the reconstruction relative $\ell_2$ for each prior distribution. The median reconstruction errors are $0.003$, $0.001$, and $0.015$, respectively, and the $95$th percentile stays below $0.04$ for all three priors. The reconstruction errors, even for the non-Gaussian prior, are relatively small. For the baseline Gaussian prior, the reconstruction error is an order of magnitude smaller than the errors in the inverse solutions reported in Table \ref{tab:latent_vs_pixel}.

\begin{table}[htbp]
\centering
\caption{The relative $\ell_2$ errors in the inverse $\bm{y}=\log \kap$ solutions obtained with the latent and full-space DPS+FNO models as functions of $\gamma$. The training, validation, and testing sets are generated from the Gaussian distribution with correlation length $l=0.2$. The inverse solutions are obtained using $N_{\mathrm{obs}}=256$ $u$ observations. The errors are averaged over 500 test samples. Also reported are the dimensions of the estimated parameter space, the total number of trainable parameters, and the time to generate one sample of the inverse solution for both DPS models. The  DPS models use the same adaptive guidance weight \cref{eq:adaptive_zeta} and FNO surrogate.}
\label{tab:latent_vs_pixel}
\small
\begin{tabular}{l r r l r r r}
\toprule
Method & Sampling dim & Params (M) & Time/sample &
  $\gamma{=}0$ & $\gamma{=}0.05$ & $\gamma{=}0.1$ \\
\midrule
L-DPS + FNO (ours) & $1024$  & 14.5 & $\sim\!1$ min & 0.022 & 0.127 & 0.178 \\
Full-space DPS + FNO    & $16384$ & 21.7 & $\sim\!5$ min & 0.023 & 0.148 & 0.229 \\
\bottomrule
\end{tabular}
\end{table}


Later in Section \ref{sec:non_gaussian}, we mix samples from three distributions, yielding a complex non-Gaussian prior. We use the combined samples to train a single latent-space DDPM model, FNO surrogate, and DPS, yielding a single L-DPS+FNO pipeline for parameter estimation without specifying any of the three underlying distributions. 

\FloatBarrier

\subsection{Effect of surrogate-based guidance}\label{sec:ablation_surrogate}

A central design choice in L-DPS is the use of a differentiable neural surrogate 
$\mathcal{F}_\phi \approx \mathcal{F}$ to provide the likelihood-gradient 
correction at each reverse-diffusion step. In the proposed method, this gradient is
computed by differentiating the data misfit through the decoder--surrogate chain in Eq \eqref{eq:chain}
rather than by repeatedly evaluating and differentiating the full PDE solver.
This surrogate-based guidance is computationally attractive, but it raises an
important question: how sensitive is the inverse reconstruction error to the
accuracy of the surrogate forward model?

To assess this dependence, we compare L-DPS reconstructions obtained with three
neural-operator surrogates: FNO, ViT, and DeepONet. In all comparisons, the VAE
encoder--decoder, latent DDPM prior, observation model, guidance rule, and
inversion hyperparameters are held fixed; only the surrogate model used to
evaluate the predicted observations and likelihood-gradient direction is varied.
This isolates the effect of the surrogate approximation on the posterior-guided
inverse solution.

We evaluate this effect in two complementary ways. First, we compare the best
validation checkpoints of the three surrogate architectures, whose forward
relative $\ell_2$ errors are reported in \cref{tab:surrogates}. Second, to
separate surrogate accuracy from architecture choice, we save intermediate
checkpoints of each surrogate during training, producing a collection of
surrogate models with different forward accuracies. Specifically, we use ten
training checkpoints for FNO and ViT and eight checkpoints for DeepONet. Each
checkpoint is inserted into the same L-DPS pipeline, and the resulting inverse
relative $\ell_2$ error is computed on the same test set. This checkpoint study
allows us to examine whether inverse accuracy is controlled primarily by the
surrogate architecture or by the forward accuracy of the surrogate.

All surrogate models are trained with the Adam optimizer and a cosine
learning-rate schedule for up to $5000$ epochs, with early stopping based on the
validation relative $\ell_2$ error using a patience of $200$ epochs; full
hyperparameters are given in \cref{tab:hp_surrogates} of Appendix \ref{app:hyper}. Unless otherwise stated, the forward errors reported below are test-set relative $\ell_2$ errors of the
best-validation checkpoint for each surrogate.

\begin{table}[htbp]
\centering
\caption{Surrogate forward error vs.\ inverse error at $N_{\mathrm{obs}} = 128$, $\gamma = 0.01$. All three surrogates share the same L-DPS pipeline (VAE + latent UNet + adaptive guidance). Forward error is the surrogate's relative $\ell_2$ test error; inverse error is the mean relative $\ell_2$ on $y = \log\kappa$ over 500 test samples.}
\label{tab:surrogates}
\small
\begin{tabular}{l r r r l}
\toprule
Surrogate & Params (M) & Forward relative $\ell_2$ & Inverse relative $\ell_2$ & Time/sample \\
\midrule
FNO       & 4.2 & 0.46\% & 0.045 & $\sim\!1$ min   \\
ViT       & 6.0 & 0.23\% & 0.042 & $\sim\!1$ min   \\
DeepONet  & 1.3 & 0.72\% & 0.051 & $\sim\!0.5$ min \\
\bottomrule
\end{tabular}
\end{table}

\begin{figure}[htbp]
\centering
\includegraphics[width=0.95\textwidth]{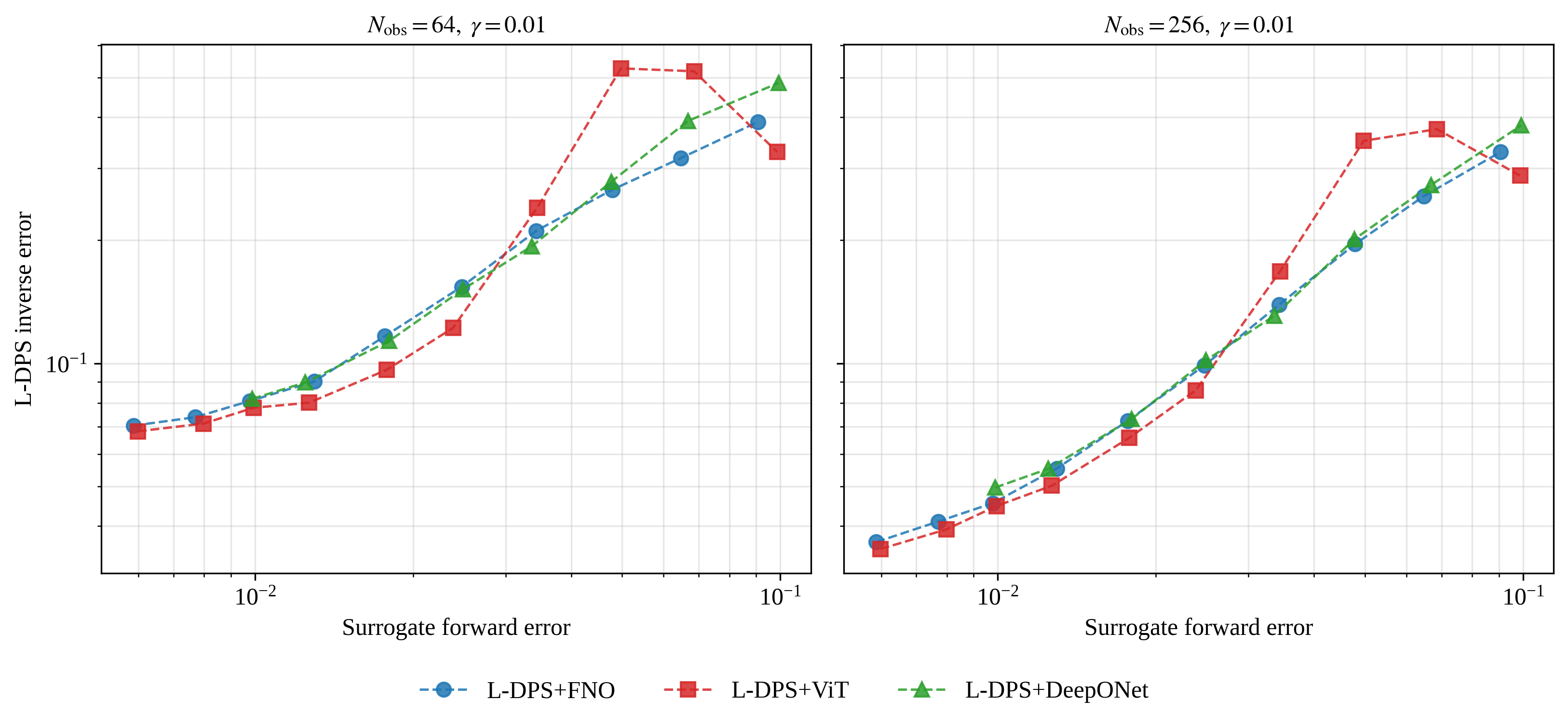}
\caption{Errors in the inverse L-DPS solutions versus errors in the FNO, ViT, and DeepONet (forward) surrogate models. The errors in the FNO and ViT surrogate models correspond to 10 training checkpoints and $8$ checkpoints for DeepONet. Left panel reports L-DPS inverse results for $N_{\mathrm{obs}}=64$ and $\gamma = 0.01$ and the right panel for $N_{\mathrm{obs}}=256$ and $\gamma = 0.01$.   
}
\label{fig:fwd_inv_scatter}
\end{figure}

Table \ref{tab:surrogates} summarizes the relative $\ell_2$ errors in the inverse solution obtained with FNO, ViT, and DeepONet surrogate models. The table also lists relative $\ell_2$ errors in the forward solutions obtained with the three surrogates. Both errors in the forward and inverse solutions are averaged over 500 test samples. 
Figure \ref{fig:fwd_inv_scatter} shows relative $\ell_2$ errors in the inverse L-DPS solutions versus errors in the FNO, ViT, and DeepONet surrogate models. The L-DPS errors are reported for $N_{\mathrm{obs}}=64$ and 256 measurements with the relative noise of $\gamma = 0.01$. 

Two patterns emerge from Table \ref{tab:surrogates} and Figure \ref{fig:fwd_inv_scatter}. First, at a fixed operating point (128 sensors, $\gamma = 0.01$), the inverse error ranks the three architectures in the same order as the forward error: ViT ($0.042$) $<$ FNO ($0.045$) $<$ DeepONet ($0.051$), matching the forward-error ordering $0.23\% < 0.46\% < 0.72\%$. Second, this ordering is not specific to the type of surrogate model but the accuracy of the surrogate model: for surrogate model errors spanning roughly an order of magnitude, the inverse error monotonically increases with the error in the forward surrogate prediction. Different surrogate models with matched forward error produce comparable inverse error.

This observation has two practical implications. First, improving the surrogate improves the inverse solution accuracy: halving the forward error translates to a proportional reduction in inverse error across the range we tested. Second, the choice of surrogate architecture is largely interchangeable in L-DPS, as long as the forward quality is matched. 

We stop short of claiming architecture independence of L-DPS in the strongest form for two reasons. The surrogate's \emph{gradient} quality is what L-DPS actually uses, and we measure only its forward solution error; at very low forward error, the gradient noise floor may differ between architectures in ways the forward solution error metric does not capture. Second, the range of forward errors we tested ($0.2\%$--$0.7\%$) is narrow relative to what could be produced by, for example, severely undertrained or poorly specified models. 

\FloatBarrier

\subsection{Effect of stabilized guidance}\label{sec:ablation_guidance}

\begin{figure}[htbp]
\centering
\includegraphics[width=\textwidth]{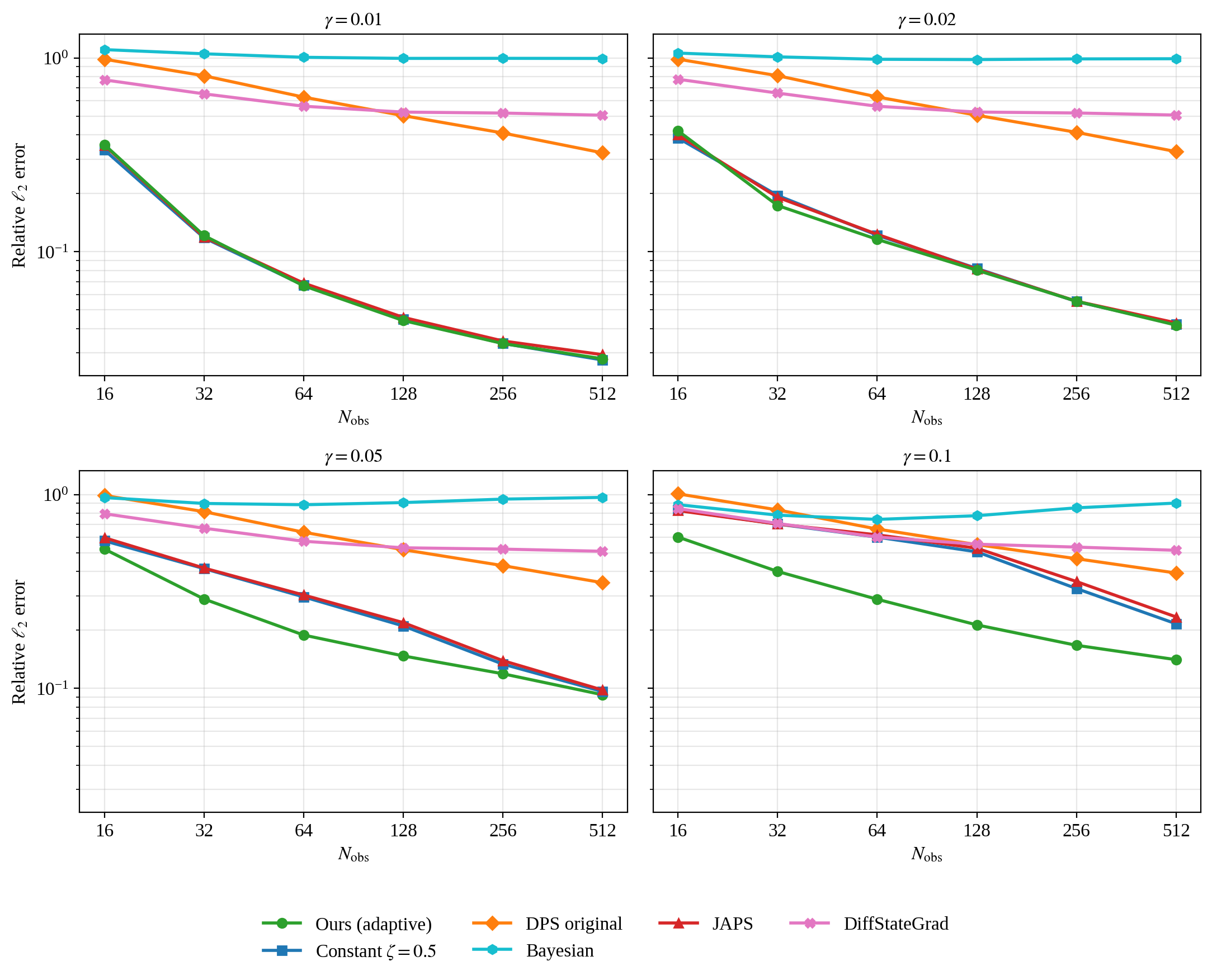}
\caption{Inverse relative $\ell_2$ errors versus the number of observations $N_{obs}$ for six guidance strategies across four noise levels $\gamma \in \{0.01, 0.02, 0.05, 0.1\}$. All methods share the same L-DPS+FNO pipeline and differ only in the guidance rule.}
\label{fig:guidance_vs_sensors}
\end{figure}

The guidance update is sensitive to two design choices: the gradient normalization (\cref{eq:stabilized_update}) and the step size $\zeta(\gamma, \eta)$. We compare six guidance strategies for the L-DPS+FNO model, including (1) the proposed here noise-adaptive rule in \cref{eq:adaptive_zeta}, (2) the constant $\zeta = 0.5$, (3) the DPS guidance proposed in \cite{chung2023diffusion}, (4) the Bayesian scaling in which $\zeta$ is set to the noise variance \cite{chung2023diffusion}, (5) JAPS  \cite{hen2025robust}, and (6) DiffStateGrad \cite{zirvi2025diffusion}. Figure \ref{fig:guidance_vs_sensors} reports the inverse relative $\ell_2$ error versus $N_{obs}$ for four noise levels $\gamma$. All methods share the same L-DPS+FNO pipeline and differ only in the guidance rule.

Three observations follow. First, our noise-adaptive rule matches or outperforms every baseline guidance rule across for all considered combinations of $N_{obs}$ and $\gamma$, and the advantage is largest at high noise: at $\gamma = 0.1$, $N_{\mathrm{obs}} = 128$ our rule gives $0.213$ versus $0.612$ for the constant-$\zeta$ baseline and $0.543$ for DPS original. Second, a single $\zeta$ cannot serve both the noiseless and noisy regimes---a large $\zeta$ overfits to noisy observations (constant $\zeta$ degrades rapidly under noise), while a small $\zeta$ underfits noiseless observations. Third, gradient normalization alone is not sufficient; JAPS and Bayesian, which also normalize the gradient, produce larger errors than our rule because they do not adapt $\zeta$ to the observation density $\eta = N_{\mathrm{obs}}/N_{\mathrm{tot}}$.

We calibrate our guidance rule \cref{eq:adaptive_zeta} on the validation set across the $(N_{\mathrm{obs}}, \gamma)$ grid, and select $(\zeta_{\max}, \zeta_{\min}, c)$ by minimizing the mean validation inverse error. The resulting values are $\zeta_{\max} = 0.5$, $\zeta_{\min} = 0.04$, and $c = 38$. With these coefficients, \cref{eq:adaptive_zeta} is within $1.5\%$ of the validation-optimal mean inverse error across the $24$ cells with $\gamma > 0$ and $N_{\mathrm{obs}} \ge 32$, while the previous hand-picked rule $\zeta(\gamma) = 0.5/(1+(\gamma/0.03)^2)$ is within $11.4\%$. The calibrated values generalize to the test set without further tuning.

\FloatBarrier

\subsection{Comparison with the KLE-MAP method}\label{sec:ablation_kle}

We compare L-DPS+FNO against the KLE-MAP baseline for the three different $\bm{y}$ priors described in Section \ref{sec:ablation_surrogate}. 
For Gaussian priors, KLE-MAP provides a natural classical baseline based on a Gaussian latent representation.
Comparison with KLE-MAP for the Gaussian prior provides a test of how effectively the latent DPS framework represents the prior structure and produces posterior-guided reconstructions.
For non-Gaussian distributions, this comparison identifies the inverse-solution errors arising from the Gaussian approximation in the KLE-MAP method. 
We instantiate KLE-MAP by retaining the $K_Y$ terms in KLE of the $\bm{y}$ covariance matrix computed from the $\bm{y}$ training samples. The number of KLE terms is selected to preserve $99.99\%$ of the variance spectrum. The MAP problem is formulated in the latent space of $y$ defined by $K_Y$ coefficients of KLE and solved using the L-BFGS algorithm. The likelihood in the MAP is computed using the FNO surrogate, which is also used in the L-DPS+FNO model.

Table \ref{tab:kle_per_prior} reports relative errors in the KLE-MAP and L-DPS+FNO inverse solutions for the three priors. The inverse solutions are obtained with $N_{\mathrm{obs}}=128$ observations with $\gamma=0.01$-noise added. We find that for all considered cases, L-DPS solutions are more accurate than KLE-MAP solutions. The gap between the two solution errors is smallest for the Gaussian prior (10\%) and largest for the binary prior, at 33\%. 

This comparison indicates that the latent diffusion model provides an effective representation of the prior structure and that the KLE Gaussian approximation can introduce significant errors, especially for the binary prior, despite using many KL terms (29 for the Gaussian prior versus 3912 for the binary prior).

\begin{table}[htbp]
\centering
\caption{ 
Relative errors in the KLE-MAP and L-DPS+FNO inverse solutions for three prior distributions of $\bm{y}$. For each prior, errors are computed for 500 test samples and reported as the mean error $\pm$ standard deviation. The inverse solutions are obtained with $N_{\mathrm{obs}}=128$ observations with $\gamma=0.01$-noise added. Also reported are the number of KLE terms, $K_Y$, and the relative difference in the KLE-MAP and L-DPS+FNO errors. 
}
\label{tab:kle_per_prior}
\small
\begin{tabular}{l r r r r}
\toprule
Prior & $K_Y$ & KLE-MAP rel.\ $\ell_2$ & L-DPS+FNO rel.\ $\ell_2$ & Gap \\
\midrule
Single-scale Gaussian (CL$=0.2$) & $29$   & $0.048\pm0.017$ & $0.044\pm0.012$ & $10\%$ \\
Multi-scale Gaussian             & $23$   & $0.045\pm0.019$ & $0.036\pm0.008$ & $26\%$ \\
Binary                           & $3912$ & $0.140\pm0.034$ & $0.105\pm0.028$ & $33\%$ \\
\bottomrule
\end{tabular}
\end{table}

\FloatBarrier

\subsection{Robustness to non-Gaussian and mixed parameter fields}\label{sec:non_gaussian}

In Section \ref{sec:ablation_kle}, we presented inverse solutions for three different parameter priors obtained with the ``problem-specific'' L-DPS+FNO models. In that section, three L-DPS+FNO models were trained, one for each prior distribution. Our results show that the errors in the inverse solutions are higher for the non-Gaussian binary prior than for the Gaussian priors with single-scale and multiscale covariances. However, we found that L-DPS+FNO solutions for the non-Gaussian prior are more accurate than the KLE-MAP solutions, where a non-Gaussian prior is approximated with a Gaussian prior. In this section, we train a foundational L-DPS+FNO model on prior samples, combining samples from the three distributions described in Section \ref{sec:ablation_kle}. Once trained, the foundational L-DPS+FNO model provides an inverse solution given measurements of $u$ without specifying which of the three priors are associated with these measurements.  

\begin{figure}[htbp]
\centering
\includegraphics[width=0.95\textwidth]{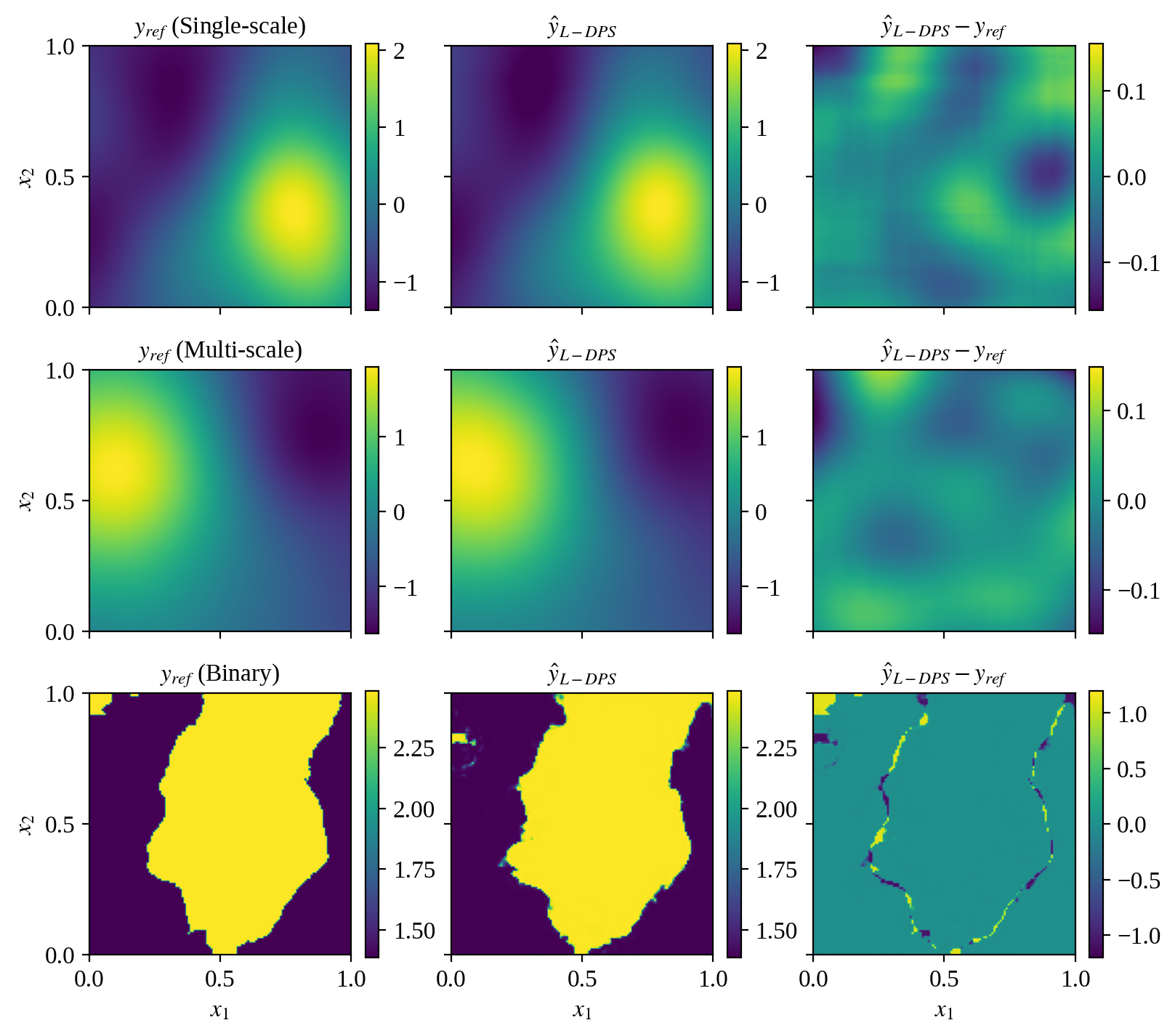}
\caption{Inverse reconstructions at $N_{\mathrm{obs}} = 128$, $\gamma = 0.01$ for three priors. Rows from top to bottom: Single-scale Gaussian, multi-scale Gaussian, Non-Gaussian binary. Columns from left to right: reference $\bm{y}_{ref}$, L-DPS+FNO estimate $\hat{\bm{y}}_{L-DPS}$, and point errors $\hat{\bm{y}}_{L-DPS}-\bm{y}_{ref}$. Solutions are shown for the test cases corresponding to the median $\ell_2$ errors.}
\label{fig:non_gaussian_inverse}
\end{figure}

\begin{figure}[htbp]
\centering
\includegraphics[width=0.95\textwidth]{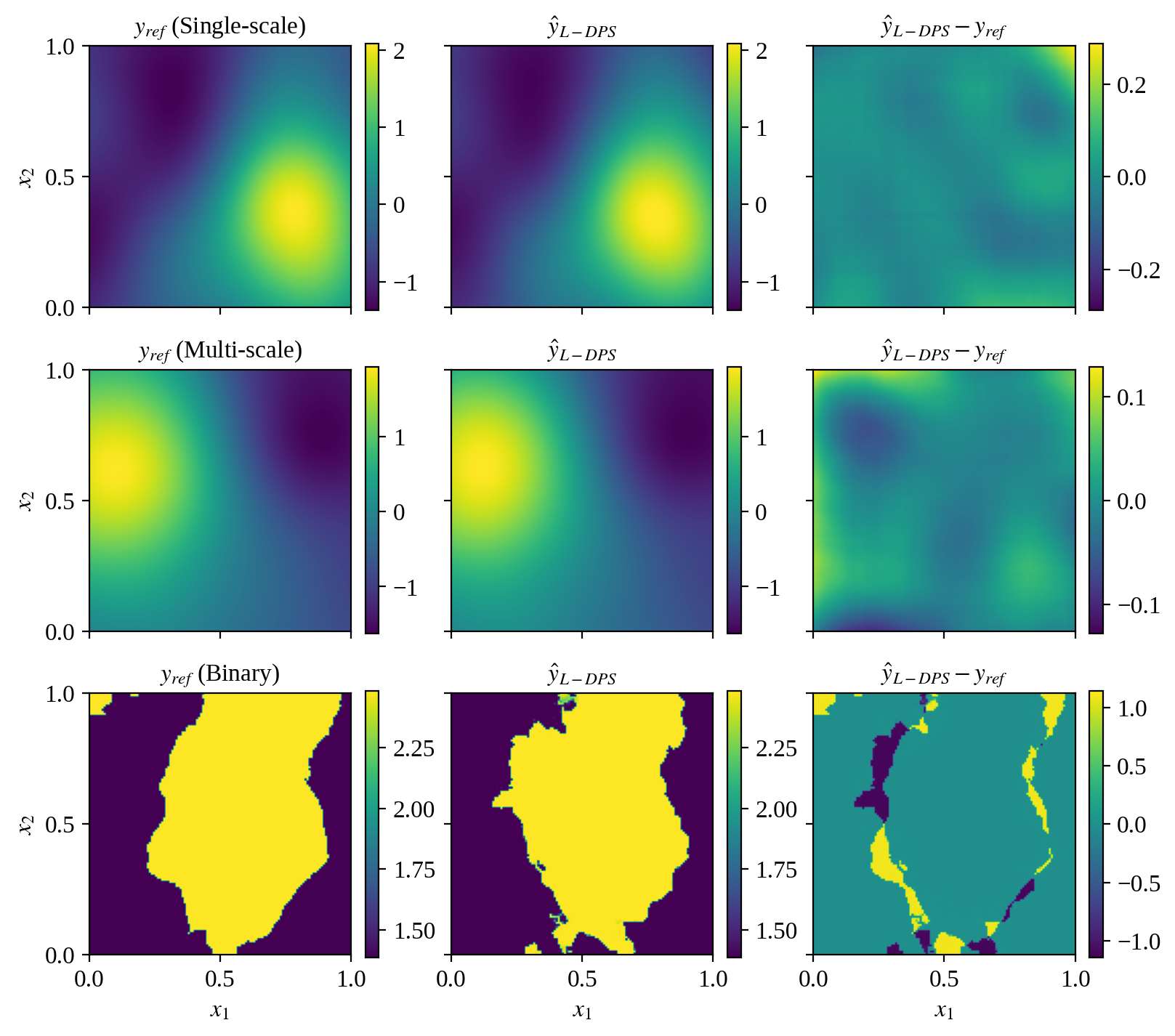}
\caption{Inverse reconstructions at $N_{\mathrm{obs}} = 128$, $\gamma = 0.01$ from a single L-DPS+FNO model trained on a mixed dataset. Rows correspond to the three source priors; the model is not told which prior a test sample came from.}
\label{fig:combined_by_source}
\end{figure}

Figure \ref{fig:non_gaussian_inverse} presents results of L-DPS+FNO models trained separately for each of the three prior distributions as described in Section \ref{sec:ablation_kle}. Specifically, Figure \ref{fig:non_gaussian_inverse} shows the reference $\bm{y}$ fields, inverse L-DPS+FNO solutions, and the point errors in the inverse solutions for the three prior distributions. The shown solutions are for the test cases with median errors across 500 test samples for each prior. Table \ref{tab:non_gaussian} reports the average relative $\ell_2$ errors over $500$ test samples. 
 Figure \ref{fig:combined_by_source} presents the inverse solutions obtained from the foundational L-DPS+FNO model obtained for the same sets of $u$ measurements (i.e., for the same test samples) as in Figure \ref{fig:non_gaussian_inverse}. Table \ref{tab:non_gaussian} summarizes average $\ell_2$ errors in the problem-specific and foundational solutions for each of the parameter distributions.

\begin{table}[htbp]
\centering
\caption{L-DPS+FNO inverse relative $\ell_2$ error at $N_{\mathrm{obs}} = 128$, $\gamma = 0.01$ across the three priors.}
\label{tab:non_gaussian}
\small
\begin{tabular}{l r r}
\toprule
Prior &  Problem-specific L-DPS+FNO & Foundational L-DPS+FNO \\
\midrule
Single-scale (baseline) & $0.044$ & $0.099$ \\
Multi-scale              & $0.036$ & $0.052$ \\
Binary                   & $0.105$ & $0.149$ \\
\bottomrule
\end{tabular}
\end{table}

Overall, the foundational L-DPS model remains competitive across the three prior families, but it does not uniformly match the accuracy of prior-specific models. 
For the multi-scale Gaussian prior, the loss of accuracy relative to the prior-specific model is modest, whereas for the single-scale Gaussian and binary priors the mixed-prior model produces larger errors. 
Nevertheless, the mixed-prior model can be applied without identifying the prior class at inference time and remains comparable to, or better than, the KLE-MAP baseline for the non-Gaussian binary case. 
These results suggest that a single mixed-prior diffusion model can provide a practical general-purpose inverse model, although prior-specific training remains preferable when the parameter-field family is known.

\FloatBarrier

\subsection{Computational efficiency}\label{sec:cost}

Table \ref{tab:cost} summarizes the per-sample inference cost (i.e., time to generate one sample of inverse solution from its posterior distribution) and the corresponding inverse error of the models considered in this work, including L-DPS+FNO, L-DPS+ViT, L-DPS+DeepONet, full-space DPS+FNO, CLDM, and InvFNO. The models are compared for the baseline setting (single-scale Gaussian parameter prior, $N_{\mathrm{obs}} = 256$, and $\gamma = 0.01$). L-DPS+FNO takes about one minute per sample and produces an inverse error of $0.022$. The full-space DPS requires about five minutes for comparable accuracy. Amortized methods are much faster at inference time: both CLDM and InvFNO achieve a solution in less than $0.071$ second per sample. However, both these methods have higher inverse solution errors. Also, amortized methods must be retrained whenever the number of measurements changes.

\begin{table}[htbp]
\centering
\caption{Inference cost and inverse error at $N_{\mathrm{obs}} = 256$, $\gamma = 0.01$. Timings are per sample computed on one A100-class GPU.}
\label{tab:cost}
\small
\begin{tabular}{l r l l}
\toprule
Method            & Inverse relative $\ell_2$ & Time/sample & Needs retraining per $N_{\mathrm{obs}}$? \\
\midrule
L-DPS+FNO (ours)  & 0.022 & $\sim\!1$ min   & No  \\
L-DPS+ViT         & 0.016 & $\sim\!1$ min   & No  \\
L-DPS+DeepONet    & 0.032 & $\sim\!0.5$ min & No  \\
Full-space DPS+FNO     & 0.023 & $\sim\!5$ min   & No  \\
CLDM              & 0.038 & $\sim\!0.07$ s  & Yes \\
Sparse InvFNO     & 0.039 & $<\!0.01$ s     & Yes \\
\bottomrule
\end{tabular}
\end{table}

\FloatBarrier

\section{Conclusions}\label{sec:conclusion}

We proposed L-DPS, an approximate Bayesian framework for posterior-guided inversion of high-dimensional coefficient fields in PDE models. The method combines a variational autoencoder to construct a compact latent representation of the parameter field, an unconditional latent diffusion model to learn an implicit prior score in this latent space, and a differentiable neural surrogate to evaluate likelihood-gradient guidance without repeated calls to the full numerical PDE solver. Posterior-guided sampling is performed directly in the learned latent space using a normalized, noise- and density-adaptive guidance update.

We evaluated L-DPS on the inverse Darcy flow problem with an unknown spatially distributed permeability field inferred from sparse and noisy pressure observations. The results show that L-DPS provides accurate and robust inverse solutions across a range of measurement counts and noise levels. Compared with full-space DPS, L-DPS achieves comparable accuracy for noiseless observations and improved accuracy under noisy observations, while reducing the per-sample inference time by approximately a factor of five. L-DPS also outperforms the amortized inverse baselines considered here, including conditional latent diffusion and inverse FNO, particularly in sparse and noisy regimes where applying likelihood guidance at inference time is beneficial.

We further compared L-DPS with the KLE-MAP baseline, in which the parameter field is represented by a Karhunen--Lo\`eve expansion and the latent prior is approximated as Gaussian. L-DPS gives lower inverse errors for all three prior classes considered. The improvement is modest for the single-scale Gaussian prior, where KLE-MAP is well matched to the prior structure, but becomes larger for the multi-scale Gaussian and binary priors. These results indicate that learning the latent prior score with a diffusion model can improve posterior-guided inversion beyond what a Gaussian latent-prior approximation can.

We also assessed the impact of surrogate model accuracy by comparing FNO, ViT, and DeepONet surrogates within the same L-DPS pipeline. The inverse error increases monotonically with the forward error of the surrogate, while different surrogate architectures with comparable forward accuracy lead to comparable inverse accuracy. This indicates that, within the range of models tested, the quality of the surrogate approximation is more important than the surrogate architecture itself.

Finally, we investigated a mixed-prior, or foundational, L-DPS model trained on samples from multiple parameter-field families. This model can be applied without specifying which prior class generated a given test case. Although prior-specific L-DPS models remain more accurate when the test prior is known, the mixed-prior model retains competitive accuracy across Gaussian and binary prior classes. This suggests that a single latent diffusion prior trained on heterogeneous parameter samples can provide a practical general-purpose inverse model, reducing the need to train a separate diffusion prior for each parameter distribution.

Several limitations remain. The numerical study focuses on single-sample posterior-guided reconstruction rather than full posterior uncertainty quantification, and all reported inversions use \(S=1\). In addition, the accuracy of L-DPS depends on the quality of the VAE representation and the differentiable surrogate, and the posterior guidance remains approximate because it uses the Tweedie plug-in likelihood, surrogate replacement of the PDE solver, and finite-step stabilized reverse diffusion. Extending the framework to larger posterior ensembles, time-dependent multiphysics models, and adaptive surrogate refinement is a natural direction for future work.

\section{Acknowledgements}

This research was supported by CUSSP (Center for Understanding Subsurface Signals and Permeability), an Energy Earthshot Research Center funded by the U.S. Department of Energy (DOE), Office of Science under FWP 81834, and the SRI (Strategic Research Initiative) Program at the UIUC’s Grainger College of Engineering.

\bibliographystyle{unsrtnat}
\bibliography{references}

@article{zong2025vae,
  title={VAE-DNN: Energy-Efficient Trainable-by-Parts Surrogate Model For Parametric Partial Differential Equations},
  author={Zong, Yifei and Tartakovsky, Alexandre M},
  journal={arXiv preprint arXiv:2508.03839},
  year={2025}
}

@inproceedings{ramesh2021zero,
  title={Zero-shot text-to-image generation},
  author={Ramesh, Aditya and Pavlov, Mikhail and Goh, Gabriel and Gray, Scott and Voss, Chelsea and Radford, Alec and Chen, Mark and Sutskever, Ilya},
  booktitle={International conference on machine learning},
  pages={8821--8831},
  year={2021},
  organization={Pmlr}
}

@inproceedings{esser2021taming,
  title={Taming transformers for high-resolution image synthesis},
  author={Esser, Patrick and Rombach, Robin and Ommer, Bjorn},
  booktitle={Proceedings of the IEEE/CVF conference on computer vision and pattern recognition},
  pages={12873--12883},
  year={2021}
}

@article{ho2020denoising,
  title={Denoising diffusion probabilistic models},
  author={Ho, Jonathan and Jain, Ajay and Abbeel, Pieter},
  journal={NeurIPS},
  year={2020}
}

@article{song2021score,
  title={Score-based generative modeling through stochastic differential equations},
  author={Song, Yang and others},
  journal={ICLR},
  year={2021}
}

@article{chung2023diffusion,
  title={Diffusion posterior sampling for general noisy inverse problems},
  author={Chung, Hyungjin and others},
  journal={ICLR},
  year={2023}
}

@article{chung2022diffusion,
  title={Diffusion posterior sampling for general noisy inverse problems},
  author={Chung, Hyungjin and Kim, Jeongsol and Mccann, Michael T and Klasky, Marc L and Ye, Jong Chul},
  journal={arXiv preprint arXiv:2209.14687},
  year={2022}
}

@inproceedings{rombach2022high,
  title={High-resolution image synthesis with latent diffusion models},
  author={Rombach, Robin and Blattmann, Andreas and Lorenz, Dominik and Esser, Patrick and Ommer, Bj{\"o}rn},
  booktitle={Proceedings of the IEEE/CVF conference on computer vision and pattern recognition},
  pages={10684--10695},
  year={2022}
}

@article{feng2025surgin,
  title={SURGIN: SURrogate-guided Generative INversion for subsurface multiphase flow with quantified uncertainty},
  author={Feng, Zhao and Yan, Bicheng and Zhao, Luanxiao and Shen, Xianda and Zhao, Renyu and Wang, Wenhao and Zhang, Fengshou},
  journal={arXiv preprint arXiv:2509.13189},
  year={2025}
}

@article{kawar2022denoising,
  title={Denoising diffusion restoration models},
  author={Kawar, Bahjat and Elad, Michael and Ermon, Stefano and Song, Jiaming},
  journal={Advances in neural information processing systems},
  volume={35},
  pages={23593--23606},
  year={2022}
}

@article{song2020denoising,
  title={Denoising diffusion implicit models},
  author={Song, Jiaming and Meng, Chenlin and Ermon, Stefano},
  journal={arXiv preprint arXiv:2010.02502},
  year={2020}
}

@article{hen2025robust,
  title={Robust Posterior Diffusion-based Sampling via Adaptive Guidance Scale},
  author={Hen, Liav and Tirer, Tom and Giryes, Raja and Abu-Hussein, Shady},
  journal={arXiv preprint arXiv:2511.18471},
  year={2025}
}

@inproceedings{zirvi2025diffusion,
  title={Diffusion state-guided projected gradient for inverse problems},
  author={Zirvi, Rayhan and Tolooshams, Bahareh and others},
  booktitle={International Conference on Learning Representations},
  volume={2025},
  pages={31217--31242},
  year={2025}
}

\appendix

\newpage
\section{Hyperparameters}\label{app:hyper}

Table \ref{tab:hp_vae} lists the hyperparameters of the VAE, Table \ref{tab:hp_diffusion} those of the latent UNet diffusion prior, and Table \ref{tab:hp_surrogates} those of the three surrogate architectures used in this work. All models are trained with the Adam optimizer, a cosine learning-rate schedule, and early stopping on the validation loss. Unless noted otherwise, training uses a single A100-class GPU. Inversion-time hyperparameters are summarized in Table \ref{tab:hp_inversion}.

\begin{table}[htbp]
\centering
\caption{VAE hyperparameters.}
\label{tab:hp_vae}
\small
\begin{tabular}{l r}
\toprule
Parameter                           & Value \\
\midrule
Input shape                         & $1 \times 128 \times 128$ \\
Latent shape                        & $4  \times 16  \times 16$  \\
Base channels                       & $64$ \\
$\beta$ (KL weight)                 & $10^{-6}$ \\
Parameters (M)                      & $4.0$ \\
Optimizer                           & Adam \\
Learning rate                       & $10^{-4}$ \\
Batch size                          & $32$ \\
Epochs (max)                        & $5000$ \\
Early-stopping patience             & $200$ \\
\bottomrule
\end{tabular}
\end{table}

\begin{table}[htbp]
\centering
\caption{Latent UNet diffusion prior hyperparameters.}
\label{tab:hp_diffusion}
\small
\begin{tabular}{l r}
\toprule
Parameter                           & Value \\
\midrule
Input/output shape                  & $4 \times 16 \times 16$ \\
Backbone                            & 2D UNet (attention at $8\times8$) \\
Base channels                       & $64$ \\
Parameters (M)                      & $5.3$ \\
Diffusion steps ($T$)               & $1000$ \\
Noise schedule                      & Linear, $\beta \in [10^{-4}, 2{\times}10^{-2}]$ \\
Optimizer                           & Adam \\
Learning rate                       & $2{\times}10^{-4}$ (cosine) \\
Batch size                          & $64$ \\
Epochs (max)                        & $5000$ \\
Early-stopping patience             & $250$ \\
\bottomrule
\end{tabular}
\end{table}

\begin{table}[htbp]
\centering
\caption{Surrogate hyperparameters.}
\label{tab:hp_surrogates}
\small
\begin{tabular}{l r r r}
\toprule
Parameter                 & FNO     & ViT     & DeepONet \\
\midrule
Parameters (M)            & $4.2$   & $6.0$   & $1.3$ \\
Input                     & $\kappa$ field & $\kappa$ field & $\kappa$ field + query grid \\
Output                    & Pressure field & Pressure field & Pressure field \\
Modes / patches / branch width & $12$ modes & $8\times8$ patches & $128$ \\
Layers                    & $4$      & $6$      & $4$ (branch), $4$ (trunk) \\
Optimizer                 & Adam     & Adam     & Adam \\
Learning rate             & $10^{-3}$ (cosine) & $10^{-3}$ (cosine) & $10^{-3}$ (cosine) \\
Batch size                & $32$     & $32$     & $32$ \\
Epochs (max)              & $5000$   & $5000$   & $5000$ \\
Early-stopping patience   & $200$    & $200$    & $200$ \\
\bottomrule
\end{tabular}
\end{table}

\begin{table}[htbp]
\centering
\caption{Inversion-time hyperparameters.}
\label{tab:hp_inversion}
\small
\begin{tabular}{l l}
\toprule
Parameter                           & Value \\
\midrule
Reverse diffusion sampler           & DDIM \\
Number of reverse steps             & $1000$ \\
Guidance weight $\zeta(\eta,\gamma)$ & $\zeta_{\max}=0.5$, $\zeta_{\min}=0.04$, $c=38$ \\
Gradient normalization              & $\ell_2$ (global) \\
Ensemble size                       & $S=1$ (no ensembling) \\
\bottomrule
\end{tabular}
\end{table}

\end{document}